%% file: main.tex
\documentclass[journal, lettersize]{IEEEtran}
\usepackage{cite}

\usepackage{amsmath,amssymb,amsfonts}
\usepackage{graphicx}
\usepackage{textcomp}
\usepackage[dvipsnames]{xcolor}

\usepackage{mathtools}  
\usepackage{amsmath}
\usepackage{amssymb}
\usepackage{tabulary}
\usepackage{booktabs}
\usepackage[inkscapearea=page]{svg}
\usepackage{auto-pst-pdf}

\usepackage{paralist}

\usepackage{bbding}
\usepackage{pifont}
\usepackage{wasysym}
\usepackage{amssymb}
\usepackage{diagbox}

\usepackage[colorlinks=true,allcolors=blue]{hyperref}
\newcommand{\tsd}{{$360^\circ$ }}

\usepackage[top=0.60in, left=0.65in, bottom=.60in, right=0.7in]{geometry}

\usepackage[font=small]{caption}
\usepackage{subcaption}
\usepackage{booktabs}
\usepackage{bbm}

\usepackage[numbers, compress]{natbib}

\usepackage{algorithm}
\usepackage{algpseudocode}

\usepackage{mathptmx}
\usepackage{amsmath}
\usepackage{amssymb}
\usepackage{amsthm}

\theoremstyle{definition}

\DeclareMathOperator*{\argmax}{arg\,max}
\DeclareMathOperator*{\argmin}{arg\,min}
\input{notation}

\newcolumntype{P}[1]{>{\centering\arraybackslash}p{#1}}

\begin{document}

\title{
    Learning Driven Elastic Task Multi-Connectivity\\ Immersive Computing Systems    
}
{
    \author{
        \IEEEauthorblockN{Babak Badnava\IEEEauthorrefmark{1}, Jacob Chakareski\IEEEauthorrefmark{2}, Morteza Hashemi\IEEEauthorrefmark{1}} \\
        \IEEEauthorblockA{
        \IEEEauthorrefmark{1} Department of Electrical Engineering and Computer Science, University of Kansas\\
        \IEEEauthorrefmark{2} College of Computing, New Jersey Institute of Technology
        }
    }
}

\maketitle

\begin{abstract}
In virtual reality (VR) environments, computational tasks exhibit an elastic nature, meaning they can dynamically adjust based on various user and system constraints. This elasticity is essential for maintaining immersive experiences; however, it also introduces challenges for communication and computing in VR systems.
In this paper, we investigate elastic task offloading for multi-user edge-computing-enabled VR systems with multi-connectivity, aiming to maximize the computational energy-efficiency (\ie computational throughput per unit of energy consumed).
To balance the induced communication, computation, energy consumption, and quality of experience trade-offs due to the elasticity of VR tasks, we formulate a constrained stochastic
computational energy-efficiency optimization problem that integrates the multi-connectivity/multi-user action space and the elastic nature of VR computational tasks.
We formulate a centralized phasic policy gradient (CPPG) framework to solve the problem of interest online, using only prior elastic task offloading statistics (\eg energy consumption, response time, and transmission time), and task information (\ie task size and computational intensity), while observing the induced system performance (\ie energy consumption and latency).
We further extend our approach to decentralized learning by formulating an independent phasic policy gradient (IPPG) method and a decentralized shared multi-armed bandit (DSMAB) method.
We train our methods with real-world 4G, 5G, and WiGig network traces and $360^\circ$ video datasets to evaluate their performance in terms of response time, energy efficiency, scalability, and delivered quality of experience. We also provide a comprehensive analysis of task size and its effect on offloading policy and system performance.
Our analysis investigates the performance trade-offs between centralized and decentralized decision-making.
In particular, we show that while CPPG requires full observability of all users’ states, it reduces latency by 28\% and energy consumption by 78\% compared to IPPG.
\end{abstract}


\section{Introduction}\label{sec:Intro}
It is envisioned that next generation wireless networks (6G-and-Beyond) will enable an unprecedented proliferation of computationally intensive and bandwidth-hungry applications (e.g. Augmented/Virtual Reality (AR/VR), and online 3D gaming~\cite{SeaGate-2019-State}).
In particular, VR use-cases 
hold tremendous potential to advance our society and impact our daily life and the economy~\cite{ChakareskiKY:20,Chakareski-2023-Millimeter}.
Currently, AR/VR applications are becoming increasingly popular in education, training, healthcare, and gaming,
reaching a global market size of \$32.64 billion in 2024~\cite{fortune-2024-vr}.
These emerging AR/VR applications rely on significant computational resources to achieve a seamless and interactive experience while on-board computational resources remain constrained.
In particular, AR/VR applications need to augment virtual elements onto the real-world environment in real-time.
The augmentation process is composed of computationally heavy tasks such as real-time sensory data processing, simultaneous localization and mapping, depth estimation, object detection, 3D object rendering, etc. Furthermore, VR devices have limited on-board energy resources.
High computational demands, continuous sensor input, and wireless connectivity drain batteries quickly, limiting the usage time of AR/VR devices. Hence, AR/VR devices must be designed with energy-efficiency in mind.
\begin{figure}[t]
    \centering
    \includegraphics[width=\linewidth, trim= 2mm 2mm 1mm 70mm, clip=true]{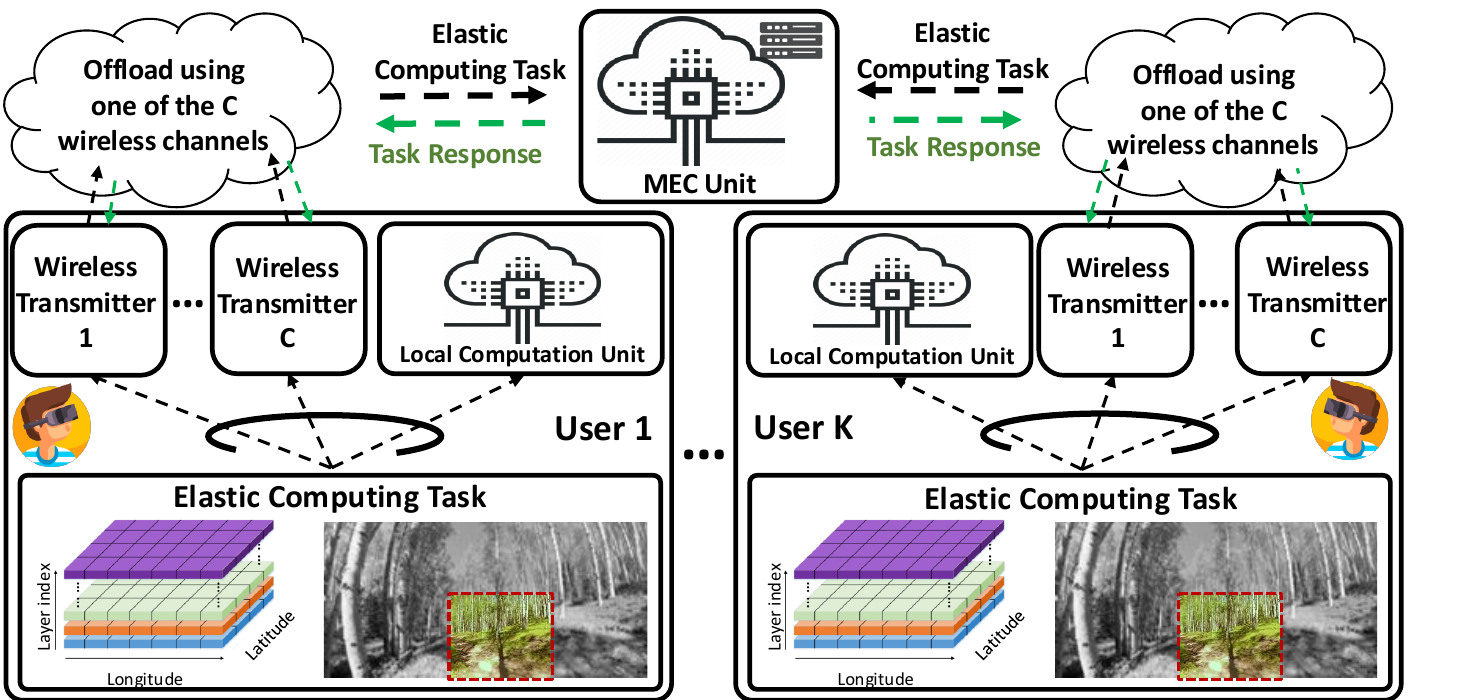}
    \caption{Multi-user multi-connectivity edge-assisted AR/VR offloading scheme: AR/VR users can compute their elastic tasks locally or offload them to a nearby MEC unit through one of the $\nbChannels$ wireless technologies available. Wireless technologies operate on different bands with unique characteristics.}
    \label{fig:sys-model}
\end{figure}

In contrast to generic computations found in other applications, AR/VR computational tasks exhibit an \emph{elastic} nature, meaning they can dynamically adjust based on system constraints, user interactions, and available resources. This elasticity is essential for maintaining immersive experiences.
For instance, VR applications can modulate rendering quality, frame rates, and latency-sensitive computations depending on network resources, edge computing availability, and computational resources. However, this adaptability comes with trade-offs. Higher fidelity graphics and lower latency enhance user experience but demand more computational power and network bandwidth, potentially overloading edge servers or causing increased energy consumption.
Conversely, reducing computational load by lowering resolution or offloading tasks to the cloud can save resources but may introduce latency, disrupting immersion. Balancing these trade-offs is crucial for ensuring smooth, high-quality VR experiences while efficiently managing system performance and resource constraints.

Given the computation and energy demands of VR systems, and to enable efficient immersive computing with elsatic tasks, we identify several distinct sets of requirements that must be satisfied:
(i) \emph{network-imposed} constraints that determine the available communication link rate~\cite{ghazikor-2024-channel},
(ii) \emph{computation-imposed} constraints that determine the available computational resources for computing VR tasks,
{(iii) \emph{elastic task-imposed} constraints that vary depending on the characteristics of tasks, required quality of experience (QoE), and available communication and computation resources.
(iv) \emph{device-imposed} constraints that determine the available computation and energy resources on the VR devices.

To integrate these requirements into a comprehensive computational framework, we propose a constrained stochastic computational energy-efficiency maximization problem for elastic AR/VR task offloading in a multi-connectivity system, as illustrated in Fig. \ref{fig:sys-model}.
In our framework, VR users can compute their elastic computational tasks locally by the AR/VR headset or offload them to a nearby multi-access edge computing (MEC) unit using one of the communication channels available to the VR headset in order to maximize computational throughput per unit of energy.
However, there are some trade-offs in this decision that need to be considered. 
On one hand, the MEC provides more computational resources, thus can process the elastic computational tasks at a higher rate, which leads to a lower computation latency and better QoE for users. Moreover, task offloading leads to less energy consumption on the AR/VR devices by the CPU/GPU.
On the other hand, computing tasks on the MEC introduces a higher communication energy consumption and a higher bandwidth requirement since tasks need to be transmitted over the air, which leads to a higher communication latency. Moreover, the AR/VR system is a stochastic system due to varying elastic task's computational and communication requirements and time-varying network conditions.

Our proposed decision-making framework considers the interplay between the variable communication and computation requirements of elastic AR/VR tasks, as well as limited on-board energy constraints.
Leveraging a state-of-the-art deep reinforcement learning (DRL) method~\cite{cobbe-2020-ppg}, and multi-agent reinforcement learning (MARL) techniques, we present three MARL agent architectures, called CPPG, IPPG, and DSMAB.
Our proposed frameworks solve a constrained stochastic multi-connectivity computational energy-efficiency maximization problem 
while incorporating user performance requirements (\ie task deadlines).
These decision-making agents learn the optimal offloading action in the VR arena by considering the communication and computation history (\ie past transmission times, past response times, and past energy consumption) and elastic task information (\ie task size, and computational intensity).
In summary, our main contributions are as follows:
\begin{enumerate}
    \item \emph{Constrained stochastic energy-efficiency maximization problem.
    }
    We introduce a multi-connectivity edge assisted AR/VR offloading schema, where AR/VR users can offload their tasks to a nearby MEC unit.
    Then, we formulate a constrained stochastic multi-connectivity computational energy-efficiency maximization problem to find the best offloading policy w.r.t. the QoE constraints, network condition, and task's computational requirements.
    \item \emph{Multi-agent computation offloading.}
    We develop a multi-agent learning-based offloading framework, in which we introduce three architectures (\ie CPPG, IPPG, and DSMAB) to find the optimal offloading policy that maximizes the energy-efficiency while meeting the QoE requirements (\ie task deadline).
    \item \emph{Multi-user task offloading simulator.}
    We develop a trace-driven gym-like~\cite{brockman2016openai} AR/VR task offloading simulator using a dataset of real-world AR/VR tasks, VR user head movement and mobility navigation information, and 4G, 5G and WiGig network traces.
    \item \emph{Extensive numerical analysis.}
    Leveraging the developed simulator, we perform an extensive numerical analysis in various system conditions.
    Our results demonstrate inherent trade-offs between centralized and decentralized decision-making processes.
    In particular, CPPG reduces the latency by 28\% and energy consumption by 78\% compared to the decentralized approach (\ie IPPG). However, it requires full observability of states of all VR users.
    Moreover, we analyze the effect of various system parameters (\eg VR computational speed, MEC computational speed, and elastic task size) on CPPG's policy and system performance (\ie response time and energy-efficiency). We show that as the task size increases, the CPPG method learns to offload more tasks to the MEC server.
    Our numerical results are representative comparison of the broader centralized versus decentralized learning-based decision-making state-of-the-art approaches.
\end{enumerate}
The rest of this paper is organized as follows. 
In Section \ref{sec:related}, we review related works and highlight our contributions.
In Section \ref{sec:system-model}, we present the multi-connectivity edge assisted AR/VR offloading system model, followed by the constrained stochastic multi-connectivity computational energy-efficiency maximization problem in Section \ref{sec:problem}.
In Section \ref{sec:solution}, we present our learning-based multi-agent decision making framework.
In Section \ref{sec:evaluation}, we provide simulation results and analysis of the performance of the proposed architectures. Section \ref{sec:conclusion} concludes the paper.

\section{Related Works}\label{sec:related}
\textbf{Generic Task Offloading:}
MEC emerged as a promising solution that provides extra computational resources near the edge of the network for latency-sensitive resource hungry applications.
However, various factors need to be considered and optimized in order to have a reliable task offloading framework.
A plethora of prior studies have investigated various aspects of task offloading services in a general setting.
A group of works investigated the problem of latency minimization~\cite{Ouyang-2019-Adaptive, Molina-2014-Joint, Wu-2021-EdgeCentric, Wang-2022-Decentralized, Yang-2022-Optimal, Huang-2019-Fine, Jia-2021-Learning}.
The works presented in \cite{Liu-2020-Joint, Zhu-2018-Cooperative, Cao-2020-UAV, Zhang-2018-Energy} investigated the energy-efficiency aspects of the task offloading services.
The authors in \cite{Jiang-2023-Joint} formulated an optimization problem that provides a guaranteed QoE while meeting long-term MEC energy constraints.
Another group \cite{Yang-2022-Multi, Sacco-2021-Sustainable, Sacco-2022-Self} investigated the trade-off between latency and energy consumption in task offloading services.
We advance these studies by accounting for unique aspects of AR/VR applications such as task elasticity, shared computational resources at the edge, and QoE considerations.





{\bf MEC-Assisted AR/VR Systems.} Integrating MEC in VR systems has been explored to provide additional: (i) computational power for decoding, rendering, and stitching of $360^\circ$ videos \cite{Chakareski-2020-6DOF, Hsu-2020-MEC} and (ii) storage space for caching $360^\circ$ video content~\cite{Dai-2020-View, Maniotis-2021-Tile,Dang-2019-Joint}.
Moreover, MEC node placement \cite{Zhang-2022-UAV,Liu-2023-On}, MEC architecture for on demand/real-time streaming \cite{Guo-2021-Design, Aung-2024-Edge,Han-2019-Real,Zhang-2023-RealVR}, communication resource management \cite{Yang-2018-Communication, Lin-2021-Resource, Guo-2020-AnAW, Chen-2019-Data}, and user scheduling \cite{Huang-2018-MAC} have been explored. Finally, MEC integrated with high-bandwidth mmWave links and multi-connectivity \cite{Chakareski-2024-Live, Ge-2017-Multipath, Liu-2019-MEC, Gupta-2023-mmWave, Ren-2019-Edge, Chakareski-2023-Millimeter} has been studied to enable multi-Gbps data rates needed for lifelike VR immersion. 
We advance these studies by formulating a dynamic decision-making framework for adaptive elastic task offloading in a multi-connectivity AR/VR system.
Our framework establishes a \emph{user-edge computing continuum}, which dynamically makes offloading decisions for elastic tasks under time-varying wireless channel condition and computational requirements. This, in turn, enhances the flexibility of the system to further improve the energy-efficiency and latency by taking into account the communication, computation, and energy requirements of elastic tasks.

\textbf{AR/VR Task Offloading:}
Among the works that integrated MEC into AR/VR application, a magnitude of works have investigated the task offloading problem for AR/VR applications to reduce the latency or improve the energy-efficiency.
The authors in \cite{Younis-2020-Latency} proposed a three-layered architecture to optimize the latency.
The authors in \cite{Song-2023-Computing} used deep reinforcement learning for offloading decision-making to reduce computational latency.
The authors in \cite{Cheng-2022-Design} propose a framework in which the VR computational task are partially offloaded to the MEC to meet the required latency for VR video streaming applications.
To tackle the energy consumption problem, the authors in \cite{Didar-2023-eAR} tried to address computation and device imposed constraints by proposing a framework for mobile AR computing to reduce energy and storage consumptions.
Moreover, the authors in \cite{Wang-2023-LEAF} proposed an edge-based energy-aware AR system that enables AR devices to dynamically change their configurations
based on user preferences
to improve the energy-efficiency of an object detection task.
The authors in \cite{Nyamtiga-2022-Edge} investigated the benefits of task offloading in terms of computational load and power consumption reduction. 
While these studies focus on alleviating the computation and device imposed constraints, our work advances these studies 
by considering the trade-offs between users' computational and communication requirements. 
Moreover, our work leverages multi-connectivity technology to further balance the communication, computation, energy, and quality of experience trade-offs arising in a networked AR/VR system.

\section{edge assisted AR/VR task offloading modeling}\label{sec:system-model}
As depicted in Fig. \ref{fig:sys-model}, we consider a multi-connectivity MEC network comprising $\nbUAVs$ AR/VR users, each equipped with a VR headset.
All headsets are equipped with computational resources (\ie CPU and GPU) to perform computational tasks arriving at the VR headsets.
Upon arrival of each elastic task (\eg depth estimation, rendering of virtual objects, \tsd video decoding), a decision-making agent selects to perform the task locally or offload it to the MEC unit using one of the $C$ communication channels available. Then, the elastic task will be computed (on the VR headset or the MEC unit) and the result will be made available to the AR/VR user.

\subsection{Computation Model}
\textbf{Task Model:}
We consider a sequence of tasks arriving at user $k$, each of which is characterized by four different features: 
\begin{inparaenum}[(i)]
    \item task size/length (in bits),  denoted by $\taskSize^k$;
    \item task computational intensity (in CPU cycles per bit), denoted by $\taskIntencity^k$;
    \item task deadline (in seconds), denoted by $\taskDeadline^k$; and
    \item task response size/length (in bits), denoted by $\resultSize^k$. 
\end{inparaenum}
Upon arrival of each task at the $k^{th}$ user, a decision making agent decides whether the task would be computed on the VR headset or offloaded to the MEC unit for computation using one of the $C$ communication channels. 
We denote $u_k \in \{0, 1, \dots C\}$ as the decision made by the decision making agent regarding the computation location.
\begin{equation}
\label{eq:action}
    u_k := 
    \begin{cases}
         u_k = 0 \;\; \Rightarrow & \text{Compute on VR headset.} \\
        u_k > 0 \;\; \Rightarrow & \text{Offload via channel $u_k$.}
    \end{cases}
\end{equation}
In case of local computation (\ie $u_k = 0$), the task is computed on the local computing unit 
of the $k^{th}$ user.
Alternatively, when the decision-making agent decides to offload the task (\ie $u_k > 0$), the task is transmitted to the MEC unit through
communication channel with index $u_k$.
Then, on the MEC side, the task is computed by the MEC unit, and the computation result is returned to the user through the same communication channel.
This leads to the following task response time:
\begin{equation}
    \taskResponse^k (S^k, I^k, u_k) =
        \begin{cases}
                u_k = 0 \;\; \Rightarrow & \taskExecution(\taskSize^k, \taskIntencity^k, \uavCPUFreq^k).  \\
                u_k > 0 \;\; \Rightarrow & 
                \taskTxTime^k(S^k, u_k) \quad \\
                & \quad + \; \taskExecution(\taskSize^k, \taskIntencity^k, \mecCPUFreq^k) \\
                & \quad + \; \resultTxTime^k(\resultSize^k, u_k).
        \end{cases} \label{eq:tr-const}
\end{equation}
Here, the function $\taskExecution(\taskSize^k, \taskIntencity^k, \uavCPUFreq^k)$ returns the task computation time given the task size $\taskSize^k$, computational intensity $\taskIntencity^k$, and VR headset's CPU frequency $\uavCPUFreq^k$. The function $\taskTxTime(S^k, u_k)$ returns the transmission time and $\resultTxTime(S^k, u_k)$ returns the task result transmission time.
Moreover, 
$\mecCPUFreq^k$ denotes CPU frequency that the MEC unit has allocated to the $k^{th}$ user.

\textbf{Task Elasticity:}
In VR environments, tasks exhibit an elastic nature, dynamically adjusting based on system constraints, user interactions, and available resources. This elasticity is evident in how video quality, computational demand, and perceived user experience interact. As illustrated in Fig. \ref{fig:task-stats}, the x-axis represents the video quality index, which corresponds to the number of enhancement layers rendered for users. The figure shows that as video quality increases, both the perceived peak signal-to-noise ratio (PSNR) and task size increase. This demonstrates the inherent trade-offs in VR systems. Higher video quality enhances user experience but comes at the cost of increased computational and network demands, potentially overloading edge servers or increasing latency. Conversely, reducing task size conserves resources but may degrade perceived quality, impacting immersion. Balancing these trade-offs is crucial to optimizing performance while maintaining an engaging VR experience.



\begin{figure}
    \centering
    \includegraphics[width=0.95\linewidth]{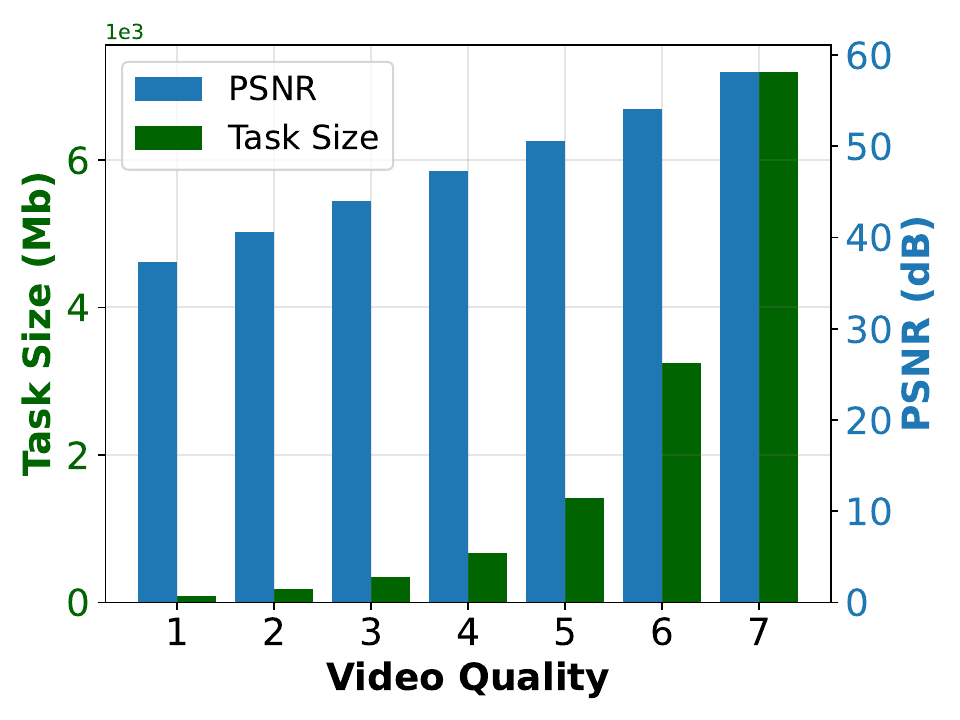}
    \caption{Task Elasticity: Task computational and communication requirements (measured via task size) and quality of experience (measured via PSNR) increase as the video quality increases, introducing challenges for VR communication and computing systems.}
    \label{fig:task-stats}
\end{figure}

\textbf{Task Computation Model:}
The task computation time depends on its size, computational intensity and the operating frequency of the CPU that computes the task.
Given a CPU with an operating frequency of $\cpuFreq$ (in cycle per second), a task with size $\taskSize$ (in bits) and computational intensity $\taskIntencity$ (in cycles per bit), the task computation time is calculated as:
\begin{align}\label{eq:cpu-exec-model}
    \taskExecution (\taskSize, \taskIntencity, \cpuFreq) = \frac{\taskSize \taskIntencity}{z(\cpuFreq)} \quad [Seconds].
\end{align}
Here, the function $z(.)$ returns the computation speed of a CPU with frequency $f$.


\textbf{Communication Model:}
The edge device transmits the task through one of the $C$ available channels to the edge server.
The expected transmission rate for a task is modeled as:
\begin{equation}
    R^k (u_k) = \frac{1}{t_e - t_s} \int_{t_s}^{t_e} R^k_t(u_k) dt,
\end{equation}
where $t_s$ and $t_e$ are the transmission start and end times, respectively, and $R^k_t(u_k)$ is the throughput provided by the uplink communication channel with index $u_k$ for the $k^{th}$ user.
The expected reception rate for a task result is modeled in a similar way over the downlink communication channel. 
Hence, the task transmission time and result reception time follow:
\begin{equation}\label{eq:tx-time}
    \taskTxTime^k(S^k, u_k) = 
    \frac{\taskSize^k}{R^k (u_k)}, \;
    \resultTxTime^k(\resultSize^k, u_k) = 
    \frac{\resultSize^k}{R^k (u_k)}.
\end{equation}
Note that although we use the same notation for uplink and downlink throughput, we employ different traces for each of them in our simulations.

\subsection{Energy Consumption Model}
The headset's battery is depleted by three different components, its CPU for local computation, its transmitter for offloading tasks to the MEC unit, and its wireless receiver for receiving results of computations.

\textbf{Computation Energy Consumption Model:}
The main source of task computation energy consumption is the CPU, and the energy consumption of the other components are negligible~\cite{Zhang-2013-Energy}.
CPU power usage consists of different factors such as dynamic power, short circuit power, and leakage power. Among all these three components, the dynamic power is the dominant part~\cite{Zhang-2013-Energy, Burd-19960-Processor}.
The dynamic power consumption is proportional to the squared supplied voltage to the chip~\cite{Burd-19960-Processor}. 
Furthermore, the clock frequency of the CPU is approximately linear proportional to the supplied voltage~\cite{Burd-19960-Processor}. 
Thus, for user $k$, the headset's CPU  energy consumption follows:
\begin{align}
\cpuEnergy^k (S^k, I^k, \uavCPUFreq^k) = \kappa \taskSize^k \taskIntencity^k (\uavCPUFreq^k)^2 \quad [Joule],
\end{align}
where $\kappa$ is the CPU capacitance factor, specific to each CPU, and $\uavCPUFreq^k$ is the headset's CPU frequency.
$\taskSize^k$ and  $\taskIntencity^k$ are task size and task computational intensity of the $k^{th}$ user, respectively.
Note that we do not consider the task computation energy consumption for the offloaded tasks since the MEC unit has unlimited power sources.

\textbf{Communication Energy Consumption Model:}
The power consumption for transmission and reception of a task over a wireless channel varies based on the wireless technology used by the headset (i.e. Wi-Fi, LTE, or 5G). 
The wireless transmitter energy consumption can be summarized by three different factors: transmission power, the efficiency of the transmit power amplifier, and a constant factor that corresponds to the power consumed by all other circuit blocks~\cite{Dusza-2013-CoPoMo, Zappone-2015-Energy}. The transmit power amplifier is usually assumed to be 100\% efficient, and the constant power consumption factor is negligible. 
Hence, the transmitter energy consumption is
modeled as:
\begin{align}\label{eq:tx-energy-consum}
    \txEnergy^k(S^k, u_k) = \taskTxTime^k(S^k, u_k) P^k_{tx}(u_k) \quad [Joule],
\end{align}
where the function $\taskTxTime^k(., .)$ returns the task transmission time over the wireless channel with index $u_k$ as described in Eq. \eqref{eq:tx-time}, and $P^k_{tx}(u_k)$ is the power allocation profile for the wireless channel with index $u_k$, according to \cite{Narayanan-2021-Variegated}. 
Similarly, the receiver energy consumption is calculated as:
\begin{align}\label{eq:rx-energy-consum}
    \rxEnergy^k(S^k, u_k) = \resultTxTime^k(S^k, u_k) P^k_{rx}(u_k) \quad [Joule],
\end{align}
where $\resultTxTime^k(S^k, u_k)$ denotes the task result transmission time over wireless channel $u_k$, and $P^k_{rx}(u_k)$ is the power consumption profile of the wireless receiver for wireless channel $u_k$. 

Note that the total task energy consumption is calculated differently for locally computed tasks and offloaded tasks.
The energy consumption of locally computed tasks consists of only the headset's CPU energy consumption,
while the energy consumption of offloaded tasks consists of transmission and reception energy consumption. 
Hence, the total energy consumption of an elastic task is obtained by:
\begin{equation}
    \totalEnergy(S^k, I^k, u_k) =
        \begin{cases}
            u_k = 0 \;\; \Rightarrow & \kappa \taskSize^k \taskIntencity^k (\uavCPUFreq^k)^2 \quad,  \\
            u_k > 0 \;\; \Rightarrow & \taskTxTime^k(S^k, u_k) P^k_{tx}(u_k) \quad \\
            & \quad  + \; \resultTxTime^k(S^k, u_k) P^k_{rx}(u_k).
        \end{cases}
\end{equation}

\section{Constrained Stochastic Multi-connectivity Energy-Efficiency Maximization}\label{sec:problem}
In this section, we formulate a multi-user multi-connectivity task offloading optimization problem.
We first examine the factors that impact the performance of such a system.
There are two key contributing factors in the overall system performance: 
(i) energy consumption $\totalEnergy^k(S^k, I^k, u_k)$
and (ii) task response time $\taskResponse^k(S^k, I^k, u_k)$.
However, both of these metrics depend on the task size and computational intensity.
Hence, we introduce computational energy-efficiency that captures all of the key contributing factor in one metric.
The computational energy-efficiency measures the computational throughput per unit of energy consumed:
\begin{align}
 \frac{
        \taskSize^k
    }{
        \taskResponse^k(S^k, I^k, u_k) \totalEnergy^k(S^k, I^k, u_k) 
    }.
\end{align}
We let $u_k$ to be the decision variable for $k^{th}$ user and formulate the optimization problem presented in Eq. \eqref{eq:ee-maximization}.
\begin{figure}[ht]
\small
\hrulefill
\begin{subequations}\label{eq:ee-maximization}
\begin{alignat}{5}
    \max_{\{u_k\}} \quad & \mathbb{E} \Big[
            \sum_{k=1}^K \frac{
                \taskSize^k
            }{
                \taskResponse^k(S^k, I^k, u_k) \totalEnergy^k(S^k, I^k, u_k) 
            } \Big]
         \quad \text{[(Bits/Second)/Joule]} \tag{\ref{eq:ee-maximization}} \\
        \text{s.t.} \quad & 
        \taskResponse^k \leq \taskDeadline^k \quad \forall \; k \in \{1,2,...,K\} \label{eq:deadline-const}\\
        \quad & 
        \taskResponse^k (S^k, I^k, u_k) =
        \begin{cases}
                u_k = 0 \;\; \Rightarrow & \taskExecution(\taskSize^k, \taskIntencity^k, \uavCPUFreq^k)  \\
                u_k > 0 \;\; \Rightarrow & 
                \taskTxTime^k(S^k, u_k) \quad \\
                & \quad + \; \taskExecution(\taskSize^k, \taskIntencity^k, \mecCPUFreq^k) \\
                & \quad + \; \resultTxTime^k(\resultSize^k, u_k) 
        \end{cases} \label{eq:tr-const} \\
        \quad & \totalEnergy(S^k, I^k, u_k) =
        \begin{cases}
            u_k = 0 \;\; \Rightarrow & \kappa \taskSize^k \taskIntencity^k (\uavCPUFreq^k)^2 \quad  \\
            u_k > 0 \;\; \Rightarrow & \taskTxTime^k(S^k, u_k) P^k_{tx}(u_k) \quad \\
            & \quad  + \; \resultTxTime^k(S^k, u_k) P^k_{rx}(u_k)
        \end{cases} \label{eq:energy-const} \\
        \quad & \sum_{k=1}^K z(\mecCPUFreq^k) < Z_{mec} \label{eq:mec-const}
\end{alignat}
\end{subequations}
\hrulefill
\end{figure}
In this optimization problem, the objective function maximizes the mean computational energy-efficiency of all VR users. 
$S^k$ and $I^k$ are two random variables corresponding to task size and task computational intensity, respectively.
Eq. \eqref{eq:deadline-const} makes sure that all users meet their task deadlines.
Eq. \eqref{eq:tr-const} defines the computation system model for each task.
Eq. \eqref{eq:energy-const} defines the energy consumption model for each task.
Finally, Eq. \eqref{eq:mec-const} enforces the maximum computation speed on the MEC unit.

There are several key challenges to solve the optimization problem in Eq. \eqref{eq:ee-maximization}, including:
\emph{(i) the Multi-User Shared Environment:} There are several users in the environment with different communication, computation, and deadline requirements.
As denoted in Eq. \eqref{eq:mec-const}, they share the MEC's computational resources among them.
This means that a change in one user's communication and computation requirements may lead to a performance degradation for other users. This is due to the fact that the MEC unit has limited computational resources that need to be allocated to a subset of users to improve the overall system performance.
\emph{(ii) the Time-Varying System Characteristics:} The environment is dynamic due to elasticity of tasks and time-varying network condition.
The amount of computational resources required to compute each elastic task varies from one task to another depending on user's request.
Moreover, each communication channel has different characteristics (\eg 5G wireless technology enables multi-Gbps transmission rates at a expense of highly dynamic channels due to blockage and mobility, while 4G wireless technology provides a less dynamic channel at a expense of a lower transmission rate).
\emph{(iii) Partially Observable Environment:} 
The optimization problem in Eq. \ref{eq:ee-maximization} needs to be solved every time a new task arrives at the headset, while some task features (\eg transmission energy consumption, transmission time, etc.) are not observable in the decision-making stage.
\emph{(iv) Elastic Tasks and Quality of Experience:} Elastic tasks have variable size and computational intensity, which effects the quality perceived by users in the environment.

Our goal is to develop a dynamic decision-making algorithm that addresses the challenges mentioned above.
Learning based decision-making methods, DRL in particular, have shown promising results in solving decision-making tasks in various applications.
The success of the learning-based methods stems from the fact that they do not rely on predefined models of the environment, but instead learn the optimal policy through repeated interactions with the environment and collecting rewards that are commensurate with the quality of the actions.
DRL methods are generally capable of capturing the uncertainty in the system due to varying task computational and communication requirements and time-varying network conditions.
However, DRL methods, in general, are not well-suited for optimization problems with inequality constraints.
Thus, we need to incorporate Eq. \eqref{eq:deadline-const} into the objective function of Eq. \eqref{eq:ee-maximization}.
Leveraging the duality principle, 
we modify the optimization problem to capture the constraint violation and penalize the DRL agent for not meeting the constraints, which leads to the Lagrangian dual problem:
\begin{equation}
\label{eq:ee-lagrange}
\small
\begin{aligned}
    \min_{\{\lambda_k\}} \max_{\{u_k\}} &  \quad
            \mathbb{E} \Bigg[
            \sum_{k=1}^K \Big[ 
            \frac{
                \taskSize^k
            }{
                \taskResponse^k(u_k) \totalEnergy^k(u_k) 
            } + \lambda_k (T_d^k - T_r^k(u_k))
            \Big]
            \Bigg] \\
    \text{s.t.} & \quad \text{Eqs.} \;\; \eqref{eq:tr-const} \text{, } \eqref{eq:energy-const} \text{, and } \eqref{eq:mec-const}
\end{aligned}
\end{equation}
which is equal to:
\begin{equation}
\label{eq:qoe-lagrange}
\small
\begin{aligned}
    \min_{\{\lambda_k\}} &  \quad
            \mathbb{E} \Bigg[
            \sum_{k=1}^K \Big[ 
            \frac{
                \taskSize^k
            }{
                \taskResponse^k(u_k^{\ast}) \totalEnergy^k(u_k^{\ast}) 
            } + \lambda_k (T_d^k - T_r^k(u_k^{\ast}))
            \Big]
            \Bigg]
            \\
    \text{s.t.} & \quad \text{Eqs.} \;\; \eqref{eq:tr-const} \text{, } \eqref{eq:energy-const} \text{, and } \eqref{eq:mec-const}
\end{aligned}
\end{equation}
Here, $u_k^{\ast}$ is the optimal offloading decision for user $k$, and $\lambda_k$ is a penalty coefficient for not meeting the deadline of the elastic task in hand, which is obtained by:
\begin{equation}
\label{eq:lambda-coeff}
\small
\begin{aligned}
    \lambda_k^{\ast} = \argmin_{\lambda_k} \;\; \underbrace{\lambda_k (T_d^k - T_r^k(u_k^{\ast}))}_{:=\mathcal{L}^{\lambda_k}}.
\end{aligned}
\end{equation}
Next, we present a multi-agent decision-making framework to address these challenges and enhance the overall performance of the system.

\section{Constrained multi-agent decision-making framework}\label{sec:solution}
To tackle the above mentioned challenges, we develop a MARL agent. 
The objective of each agent in a multi-agent setting is to maximize its expected cumulative reward:
\begin{equation}
\label{eq:return-func}
    J(\pi_k) = \mathbb{E} 
    \Bigg[
        \sum_{t=0}^{\infty} \gamma^t r(s_t, a_t) | \pi_k, \pi_{-k}
    \Bigg].
\end{equation}
Here, $\pi_k$ is the agent $k$'s policy, $\pi_{-k}$ represents the policies of all other agents, $r(s_t, a_t)$ is the reward signal received by agent $k$ at time $t$ based on the current global state $s_t$ and joint action $a_t = [u_1^t, u_2^t, ..., u_K^t]$, and $\gamma \in [0, 1)$ is the discount factor for future rewards. By letting $\gamma = 0$, the objective changes to maximizing the immediate reward, which is equivalent to multi-agent contextual MAB problem:
\begin{equation}
    J(\pi_k) = \mathbb{E} 
    \big[
        r(s_t, a_t) | \pi_k, \pi_{-k}
    \big].
\end{equation}
However, one can only set $\gamma = 0$ if there is no temporal dependency between rewards. We will show later that Eq. \eqref{eq:reward-func} fully captures the reward for a decision, which allows us to set $\gamma = 0$.
In the rest of this section, we present three frameworks for decision making in such a setting.

\subsection{Centralized Phasic Policy Gradient (CPPG)}
In this section, we present a centralized training and execution approach, called CPPG, for solving Eq. \eqref{eq:ee-lagrange}.
First, we describe the decision-making flow for each task, and then present the details of the CPPG agent (\ie state, action, reward, and architecture of the neural network). Finally, we delve into the details of the training process.

\textbf{Decision-Making Flow and State:}
At each time step, the CPPG agent observes the state of the environment $s_t = [s_1^t, s_2^t, .., s_K^t]$, which comprises the state of each agent in the environment.
The $k^{th}$ agent's state $s_k^t$ includes \emph{task information and communication and computation history}.
The task information includes task size $S_k$ and task computational intensity $I^k$.
The communication and computation history includes past transmission times, past response times, and past energy consumptions.
These metrics help the CPPG agent to learn the relationship between the task information, communication and computation history and actions that lead to the highest reward for each user.

\textbf{Action:}
Once the state $s_t$ is observed, the CPPG agent takes a joint action $a_t = [u_1^t, u_2^t, .., u_K^t]$, where
$u_k^t \in \{0, 1, ... C \}$ determines whether the $k^{th}$ agent's elastic task is computed locally or offloaded to the MEC unit using one of the $C$ communication channels as described in Eq. \eqref{eq:action}.

\textbf{Reward:}
After the joint action is applied to the environment, the state of the environment changes, and each agent receives a reward $r^k_{t+1}$.
The goal of our agent is to maximize the objective function in Eq. \eqref{eq:ee-lagrange}.
This objective function accounts for energy-efficiency, task response time, and deadline constraint.
Then, we employ the following reward function for each user:
\begin{equation} \label{eq:reward-func}
    r (s_k^t, u_k^t) =  
        \frac{
            \taskSize^k
        }{
            \taskResponse^k(u_k^t) \totalEnergy^k(u_k^t) 
        } + \lambda_k (T_d^k - T_r^k(u_k^t))
\end{equation}
This reward function does not have any temporal dependency and captures the reward of an elastic task completely, which allows us to set the $\gamma = 0$ in Eq. \eqref{eq:return-func}.

\textbf{CPPG Architecture:}
The CPPG agent, as shown in Fig. \ref{fig:cppg-arch}, is composed of an actor network $\omega$ and a critic network $\omega_v$.
The actor network outputs policies of all agents along with a set of auxiliary values that estimate the state values for each of the agents separately.
The critic network outputs the estimated state value for each of the agents separately.
This architecture enables CPPG to capture the effect of shared computational resources on the reward each user receives and account for them during the training process.
Moreover, this architecture reduces the interference between policy and value loss, while distilling features from the value function into the policy network \cite{cobbe-2020-ppg}.
Both actor and critic networks are composed of a convolutional layer and a dense layer. We employ a three-phase training procedure, consisting of a policy training phase, an auxiliary training phase~\cite{cobbe-2020-ppg, Deheng-2020-Mastering, Tianchi-2023-Buffer}, and a deadline coefficient optimization phase.

\begin{figure}
    \centering
    \includegraphics[width=0.95\linewidth, trim= 0mm 45mm 65mm 0mm, clip=true]{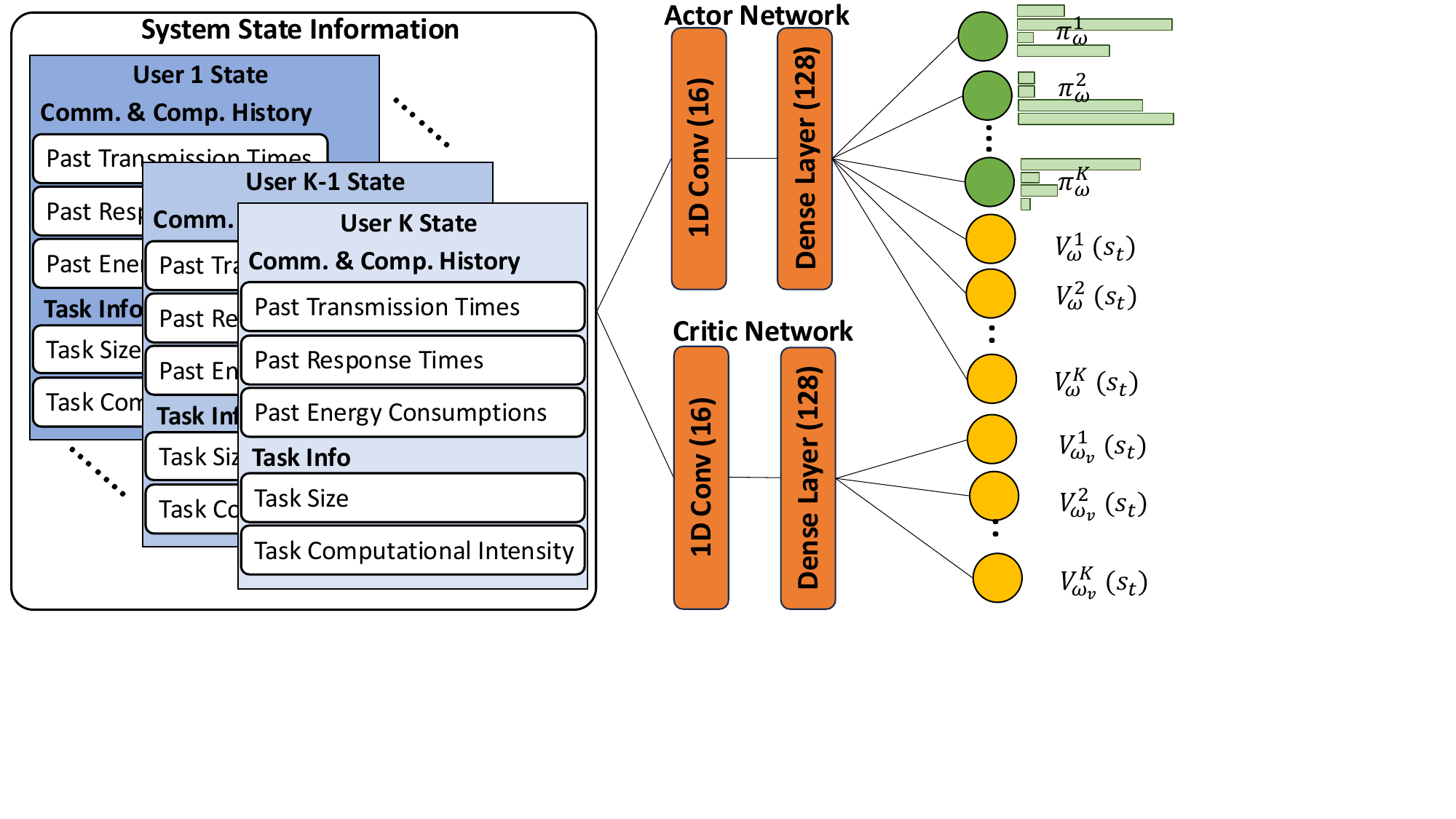}
    \caption{The CPPG agent follows an actor-critic model. The actor makes a joint offloading decision for all the users. The critic estimates the state value  for each of the users in the environment.}
    \label{fig:cppg-arch}
\end{figure}

\textbf{Policy Optimization:}
In the policy training phase, we update the actor and critic networks.
The policy network is trained by dual-clip proximal policy optimization (PPO) \cite{schulman-2017-proximal}:
\begin{equation}
\label{eq:d-clip-loss}
    \begin{aligned}
        \mathcal{L}^{DClip} = \hat{\mathbb{E}}
            \Big[ \indicator(\hat{A}_t < 0)\max(\mathcal{L}^{PPO}, c\hat{A}_t) + 
            \indicator(\hat{A}_t \geq 0)\mathcal{L}^{PPO} \Big].
    \end{aligned}
\end{equation}
Here, $\indicator(.)$ is a binary indicator function and $\hat{A}_t = r_t + \gamma V_{\omega_v}(s_{t+1}) - V_{\omega_v}(s_t)$ is the advantage function calculated based on the current state value estimate and the discount factor $\gamma = 0$, and $\mathcal{L}^{PPO}$ is the surrogate vanilla PPO loss:
\begin{equation}
\label{eq:ppo-loss}
    \begin{aligned}
        \mathcal{L}^{PPO} = \mathcal{L}^{Clip}(\pi_{\omega}, \hat{A}_t) + \beta H_{\omega}(s_t) + \mathcal{L}^{Value},
    \end{aligned}
\end{equation}
where $H_{\omega}(s_t)$ is the entropy of all the policies and $\beta$ is the entropy weight.
The entropy loss and its associated weight balance the trade-off between exploration and exploitation in the learning process.
$\mathcal{L}^{Value}$ is the value network loss:
\begin{equation}
\label{eq:value-loss}
    \begin{aligned}
        \mathcal{L}^{Value} &= \mathbb{E}\left[ 
            \frac{1}{2}
            (V_{\omega_v}(s_t) - V_{targ}(s_t))^2 
        \right],
    \end{aligned}
\end{equation}
and $\mathcal{L}^{Clip}$ is the single-clip policy loss defined as: 
\begin{equation}
\label{eq:clip-loss}
    \begin{aligned}
        \mathcal{L}^{Clip} = \min \Big[ &\rho(\pi_{\omega}, \pi_{\omega_{old}}) \odot \hat{A}_t, \\ 
 & clip(\rho(\pi_{\omega}, \pi_{\omega_{old}}), 1 \pm \varepsilon) \odot \hat{A}_t \Big].
    \end{aligned}
\end{equation}
Here $\odot$ denotes element-wise multiplication, and $\rho(\pi_{\omega}, \pi_{\omega_{old}}) \odot \hat{A}_t$ measures the gained/lost reward as the policy for each agent changes, which determines how a change in one agent's policy affects other agents reward, (\ie energy-efficiency and response time).
The change in the agents' policies w.r.t. their old policy is measured via 
$\rho(\pi_{\omega}, \pi_{\omega_{old}}) = \left[\rho_1, \rho_2, \dotsb, \rho_K \right]$, which is a vector of size $K$.
Each element of this vector $\rho_k$ measures the changes for one agent in the environment (\ie the joint probability ratio of the new policy to the old policy):
\begin{equation}
\label{eq:rho-i}
    \begin{aligned}
        \rho_k =
        \frac{
            \pi_{\omega} (u^t_k|s_t)
        }{
            \pi_{\omega_{old}} (u^t_k|s_t)
        }.
    \end{aligned}
\end{equation}

\textbf{Auxiliary Training Phase:}
In the auxiliary phase, we further optimize the actor and critic networks according to a joint loss function $\mathcal{L}^{Joint}$ on all the experiences.
The joint loss function $\mathcal{L}^{Joint}$ is composed of a behavioral cloning loss and an auxiliary value loss:
\begin{equation}
\label{eq:joint-loss}
    \begin{aligned}
        \mathcal{L}^{Jonit} = 
        \mathbb{E}
        \Bigg[ 
            \sum_{k=1}^K 
            \mathbf{KL} \left(\pi^k_{\omega_{old}} (s_t), \pi^k_{\omega} (s_t)\right) 
        \Bigg] 
        + \mathcal{L}^{Aux}.
    \end{aligned}
\end{equation}
Here, $\mathbf{KL}(.,.)$ is the behavioral cloning loss, representing the KL-divergence between the original policy and the updated policy of each of the agents.
The $\mathcal{L}^{Aux}$ in Eq. \eqref{eq:joint-loss} updates the auxiliary value by minimizing the mean squared loss function:
\begin{equation}
\label{eq:aux-loss}
    \begin{aligned}
        \mathcal{L}^{Aux} &= \mathbb{E}\left[ 
            \frac{1}{2}
            (V_{\omega}(s_t) - V_{targ}(s_t))^2 
        \right].
    \end{aligned}
\end{equation}

\textbf{Deadline Coefficient Training Phase:}
In the deadline coefficient training phase, we update the deadline coefficients $\lambda_k$ for each of the agents by solving Eq. \eqref{eq:lambda-coeff}.
Each $\lambda_k$ is updated to increase if the response time is higher than the target deadline $T_d^k$, which leads to the following sub-gradient update rule for the deadline coefficient:
\begin{equation}\label{eq:coeff-update}
    \lambda_k \leftarrow \lambda_k + \eta \nabla_{\lambda_k}\mathcal{L}^{\lambda_k}(u_k).
\end{equation}
This modification in the optimization problem effectively handles the reward magnitude change during the training phase.

\textbf{Training Algorithm:}
Algorithm~\ref{alg:cppg-alg} presents the training process of the CPPG agent.
The training continues for multiple iterations until convergence.
Each iteration consists of four phases.
In the first phase, we perform the current policy $\pi_{\omega}$ on a randomized environment to collect new experiences (lines 3-7).
We first sample the joint action, and then apply the chosen action on the edge-assisted AR/VR task offloading environment (see $CentralizedPolicy$ in Alg. \ref{alg:cppg-alg}).
Then, we store the resulting transition in a replay buffer for training purposes.
In the second phase (\ie policy training phase), we update both the actor and critic networks.
We use a random batch from the replay buffer to compute the dual-clip PPO loss $\mathcal{L}^{DClip}$, and update the networks (lines 9-11).
In the third phase (\ie auxiliary training phase), we further update the actor and critic networks by optimizing the behavioral cloning and value losses using all the replay buffer data (lines 12-15).
Finally, in the last phase, we update the deadline coefficients 
using the loss functions defined for the coefficients in Eq. \eqref{eq:ee-lagrange} (lines 16-18).


\begin{algorithm}[t]
\caption{CPPG \& IPPG Training Process}
\label{alg:cppg-alg}
\begin{algorithmic}[1]
    \Require $alg\_name \in \{CPPG,\; IPPG \}$
    \State Initialize $\mathcal{B} \gets \varnothing$
    \For{task $=1,2,\dots$}
        \If{$alg\_name = CPPG$}
            \State \Call{CentralizedPolicy}{$s_t, \mathcal{B}$}
        \Else
            \State \Call{DeCentralizedPolicy}{$s_t, \mathcal{B}$}
        \EndIf
        \If{task $\% \; N_{update} = 0$}
            \For{$i = 1,2,\dots, N_{Policy}$}  \Comment{Policy training phase}
                \State Optimize $\mathcal{L}^{DClip}$ w.r.t. $\omega, \omega_{v}$ 
            \EndFor
            \For{$i = 1,2,\dots, N_{aux}$} \Comment{Auxiliary training phase}
                \State Optimize $\mathcal{L}^{Value}$ w.r.t. $\omega_v$ 
                \State Optimize $\mathcal{L}^{Joint}$ w.r.t. $\omega$ 
            \EndFor
            \For{$i = 1,2,\dots, N_{\lambda}$} \Comment{Coefficient training phase}
                \State $\lambda_k \gets \lambda_k + \eta \nabla_{\lambda_k}\mathcal{L}^{\lambda_k}(u_k) \quad \forall k = 1 \dots K$ 
            \EndFor
            \State $\mathcal{B} \gets \varnothing$ \Comment{Clear replay buffer}
            \State $\omega_{old} \gets \omega$, $\omega_{v_{old}} \gets \omega_v$ \Comment{Update target network}
        \EndIf
    \EndFor

    \Procedure{CentralizedPolicy}{$s_t, \mathcal{B}$}
        \State $a_t \sim \pi_{\omega} (a_t|s_t)$ \Comment{Joint offloading action}
        \State $s_{t+1}, r_{t+1} \gets \text{ACT}(s_t, a_t)$ 
        \State $\mathcal{B} \gets \mathcal{B} \cup \{s_t, u_t, r_{t+1}, s_{t+1}, V_{targ}(s_t)\}$ 
    \EndProcedure
    
    \Procedure{DeCentralizedPolicy}{$s_t, \mathcal{B}$}
        \For{$k=1...K$}
            \State $u_k^t \sim \pi_{\omega}^k (u_k^t|s_k^t)$ \Comment{User $k$ offloading action}
        \EndFor
        \State $a_t = [u_1^t, u_2^t, ..., u_K^t]$
        \State $s_{t+1}, r_{t+1} \gets \text{ACT}(s_t, a_t)$
        \For{$k=1...K$}
            \State $\mathcal{B} \gets \mathcal{B} \cup \{s_k^t, u_t, r^k_{t+1}, s^k_{t+1}, V^k_{targ}(s_k^t)\}$ 
        \EndFor
    \EndProcedure
\end{algorithmic}
\end{algorithm}

    

\begin{figure}
    \centering
    \includegraphics[width=0.95\linewidth, trim= 55mm 85mm 85mm 55mm, clip=true]{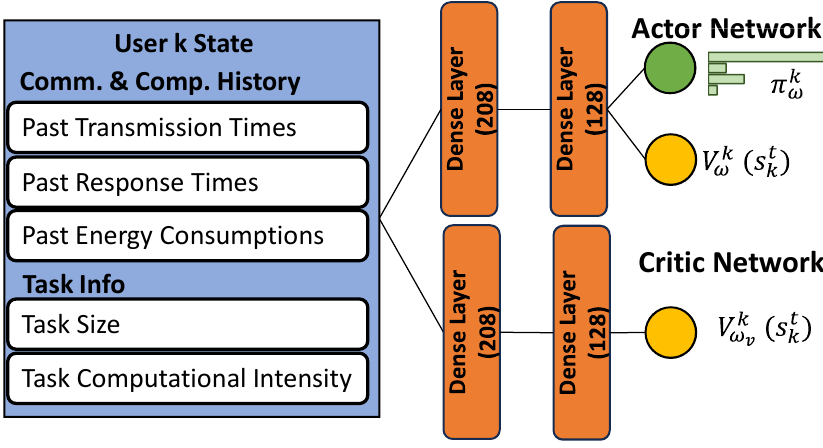}
    \caption{The IPPG agent observes the state of each agent and makes an offloading decision for the newly arrived elastic task.}
    \label{fig:dppg-arch}
\end{figure}

\subsection{Independent Phasic Policy Gradient}\label{sec:IPPG}
To extend the CPPG framework to a decentralized setting, we present the independent phasic policy gradient (IPPG) method to solve Eq. \eqref{eq:ee-lagrange}.
Similarly, the IPPG is composed of an actor network $\omega$ and a critic network $\omega_v$. 
However, unlike CPPG, both actor and critic only observe the state of one user, $s_k^t$, at a time.
Moreover, the actor learns the policy, $\pi_{\omega}^k$, and an auxiliary value function $V_{\omega}^k(s_k^t)$ of one user, while the critic learns the value function $V_{\omega_v}^k(s_k^t)$ of one user as shown in Fig. \ref{fig:dppg-arch}.
This architecture, unlike CPPG, does not depend on the number of users in the system.
Moreover, this architecture enables us to deploy the actor on each user after training, which in turn enables decentralized decision-making\cite{Yu-2022-PPO}.

\textbf{Policy Optimization:}
Similar to CPPG, we employ a three-phase training procedure for IPPG agent. 
However, the change in the architecture leads to a few changes in the loss functions.
The element-wise multiplication in the single-clip policy loss defined in Eq. \eqref{eq:clip-loss} is replaced with a scaler multiplication, which leads to:
\begin{equation}
\label{eq:clip-loss-mod}
    \begin{aligned}
        \mathcal{L}^{Clip} = \min \Big[ \rho(\pi_{\omega}, \pi_{\omega_{old}}) \hat{A}_t,\; clip(\rho(\pi_{\omega}, \pi_{\omega_{old}}), 1 \pm \varepsilon) \hat{A}_t \Big].
    \end{aligned}
\end{equation}
This means that 
$\rho(\pi_{\omega}, \pi_{\omega_{old}})$
is also a scaler, and measures the changes in policy of the $k^{th}$ agent:
\begin{equation}
\rho(\pi_{\omega}, \pi_{\omega_{old}}) = 
\frac{
    \pi_{\omega} (u^t_k|s_k^t)
}{
    \pi_{\omega_{old}} (u^t_k|s_k^t)
}
\end{equation}
Moreover, the critic network loss is updated to learn the state-value function for one user:
\begin{equation}
\label{eq:value-loss-mod}
    \begin{aligned}
        \mathcal{L}^{Value} &= \mathbb{E}\left[ 
            \frac{1}{2}
            (V^k_{\omega_v}(s_k^t) - V^k_{targ}(s_k^t))^2 
        \right].
    \end{aligned}
\end{equation}

In the axillary training phase, the behavioral cloning loss in Eq. \eqref{eq:joint-loss} changes to clone behavior of one single agent at a time, which leads to the following joint loss function:
\begin{equation}
\label{eq:joint-loss-mod}
    \begin{aligned}
        \mathcal{L}^{Jonit} = 
        \mathbb{E}
        \Bigg[ 
            \mathbf{KL} \left(\pi^k_{\omega_{old}} (s_k^t), \pi^k_{\omega} (s_k^t)\right) 
        \Bigg] 
        + \mathcal{L}^{Aux}_k.
    \end{aligned}
\end{equation}
Here $\mathcal{L}^{Aux}_k$ is the auxiliary value loss function of a single user:
\begin{equation}
\label{eq:aux-loss-mod}
    \begin{aligned}
        \mathcal{L}^{Aux}_k &= \mathbb{E}\left[ 
            \frac{1}{2}
            (V_{\omega}^k(s_k^t) - V_{targ}^k(s_k^t))^2 
        \right].
    \end{aligned}
\end{equation}
The deadline coefficient training phase stays the same as described in the previous section.

\textbf{Training Methodology:}
Similar to CPPG, the IPPG training continues for multiple iterations until convergence, and each iteration consists of three phases.
However, the first phase (\ie policy execution phase) is different due to differences between architecture of CPPG and IPPG.
The IPPG observes the state of each user separately (in $K$ passes) and then makes offloading decisions for each user.
Moreover, we employ parameter sharing~\cite{Justin-2020-Parameter} during the training to enhance the sample efficiency and scalability of the IPPG.
The rest of the training is similar to CPPG training. 
In the policy training phase, the dual-clip PPO loss $\mathcal{L}^{DClip}$ with some modification to Eq. \eqref{eq:clip-loss-mod} is optimized.
In the auxiliary training phase, $\mathcal{L}^{Value}$ (see Eq. \eqref{eq:value-loss-mod}) and $\mathcal{L}^{Joint}$ (see Eq. \eqref{eq:joint-loss-mod}) are optimized.
Finally, the deadline coefficient training phase is performed and the deadline coefficients $\lambda_k$ for each user is updated using Eq. \eqref{eq:coeff-update}.






\subsection{Decentralized Shared Multi-Armed Bandit}
In this section, we formulate a decentralized shared multi-armed bandit (DSMAB) framework to solve Eq. \eqref{eq:ee-lagrange}.
Similar to IPPG, at each time step the DSMAB agent observes the state of the environment $s_k^t$. 
However, the state information is collected differently.
Here, the state of the environment comprises the contextual information of each of the actions $s_k^t = [c_0^k, c_1^k, .., c_C^k]$.
The $c^{th}$ action's contextual information includes \emph{task information} and \emph{computation history} of action $c$.
The task information includes the task size $s_k^t$ and computational intensity $I^k_t$.
The computation history includes past response times and past energy consumptions of each action.
These metrics enable the DSMAB agent to learn the relationship between task characteristics, computational history, and the actions that yield the highest user rewards.
MAB methods optimize policy by balancing exploration and exploitation trade-off, ensuring that the agent sufficiently explores the action space to identify high-reward actions.
Various exploration strategies exist, such as Upper Confidence Bound (UCB), Thompson Sampling (TS), and epsilon-greedy, each offering different trade-offs between exploration and exploitation.
DSMAB introduce a unified MAB architecture that integrates these exploration strategies into a single cohesive approach, allowing for adaptive and efficient decision-making.

\textbf{DSMAB Architecture:}
The DSMAB agent, as shown in Fig. \ref{fig:imab-arch}, approximates the context-action value function a linear model, a Gaussian process, or a neural network. 
This approximation is then passed to a policy layer, which defines a probability distribution over the actions based on the estimated values.
The policy layer employs various exploration strategies, such as $\epsilon$-greedy, UCB, and Thompson Sampling, to select actions based on the estimated values.


\begin{figure}
    \centering
    \includegraphics[width=0.95\linewidth, trim= 0mm 63mm 54mm 0mm, clip=true]{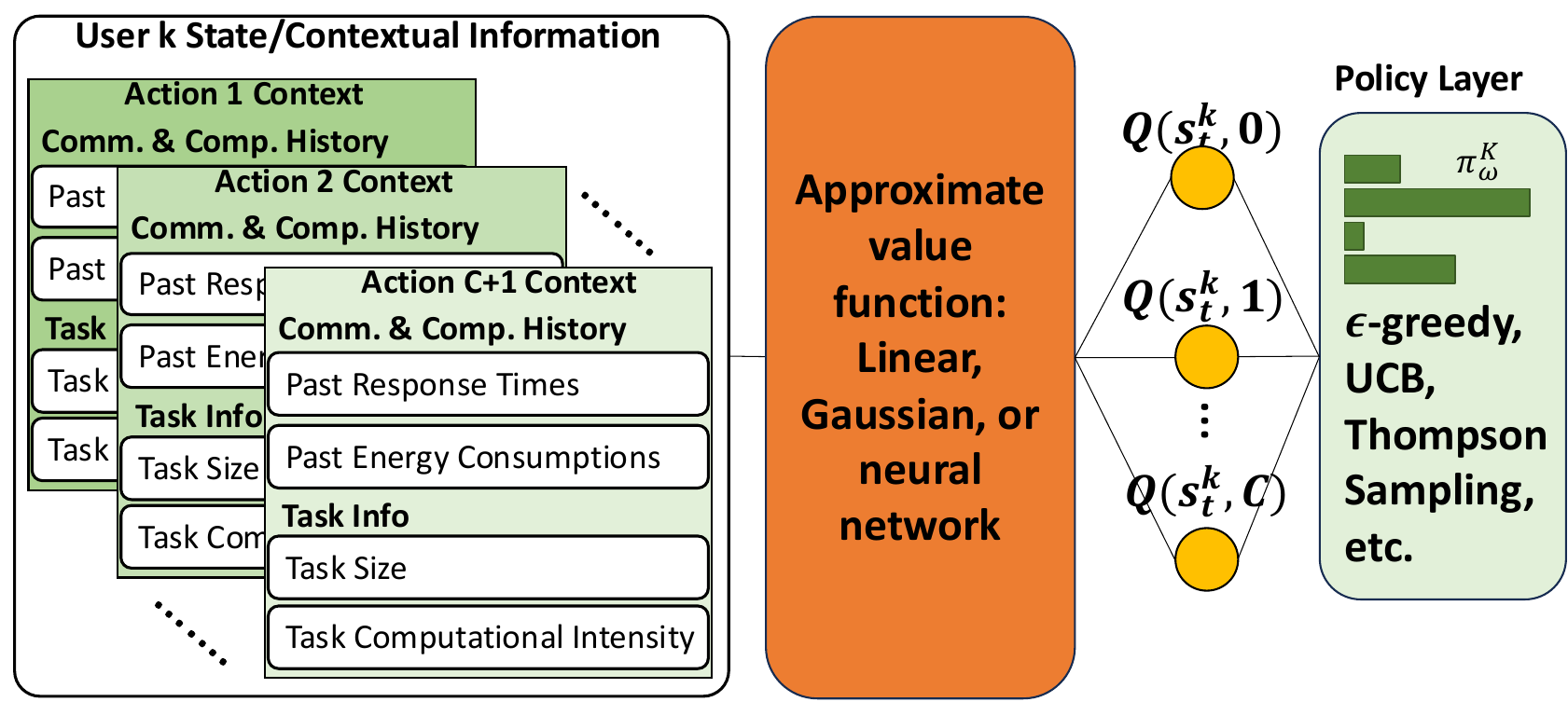}
    \caption{DSMAB agent observes contextual information of a user's actions, estimates the value of performing each action, and based on the estimation makes an offloading decision for the elastic task.}
    \label{fig:imab-arch}
\end{figure}

\textbf{Neural $\epsilon$-Greedy:}
The Neural $\epsilon$-Greedy agent, referred to as INN $\epsilon$-Greedy, employs a two layer neural network to estimate the action values by minimizing the following loss function:
\begin{equation}
\label{eq:dec-value-loss}
    \begin{aligned}
        \mathcal{L}^{Dec.}(\theta) &= \mathbb{E}\left[ 
            \frac{1}{2}
            (r_t^k - Q(s_k^t, u^t_k; \theta))^2 
        \right] + ||\theta||^2_2.
    \end{aligned}
\end{equation}
Here the first term computes the mean squared error (MSE) between the predicted reward for an action and the received reward given the neural network parameters $\theta$.
The second term regularizes the network parameters to prevent overfitting and an exploding gradient problem during the training process.
The policy layer follows an $\epsilon$-Greedy strategy, selecting a random action with probability $\epsilon$ and the best action with probability $1-\epsilon$:
\begin{equation}
    u^t_k = \argmax_{u} Q(s_k^t, u; \theta)
\end{equation}

\textbf{Neural-UCB:}
The Neural-UCB agent, referred to as INN-UCB, employs a two layer neural network and the loss function presented in Eq. \eqref{eq:dec-value-loss} to estimate action values. Unlike INN $\epsilon$-Greedy, which explores randomly, INN-UCB balances exploration and exploitation using upper confidence bound (UCB) algorithm. This results in the following action selection mechanism in the policy layer:
\begin{equation}
    u^t_k = \argmax_{u} Q(s_k^t, u; \theta) + \sqrt{g(s_k^t; \theta)^T Z^{-1}g(s_k^t; \theta)}
\end{equation}
Here, $g(s_k^t; \theta)$ represents the feature vector from the last layer of the neural network, and $Z$ is the regularized feature covariance matrix, given by: 
\begin{equation}
\label{eq:ucb-bound}
    Z = \mu I + g(s_k^t; \theta)g(s_k^t; \theta)^T
\end{equation}
The positive regularization term $\mu$ ensures that $Z$ remains invertible~\cite{Zhou-2020-Neural}. 
We also employ a linear version of this method, refereed to as ILin-UCB, in which we use a linear function approximator to estimate the action values.

\textbf{Neural Thompson Sampling:}
The Neural Thompson Sampling agent, referred to as INN-TS, follows a similar structure, employing a two-layer neural network and the loss function from Eq. \eqref{eq:dec-value-loss} to estimate action values.
However, unlike UCB, which explicitly computes confidence bounds, Thompson Sampling handles exploration by maintaining a Gaussian posterior distribution over action values.
At each step, the action values are drawn from a normal distribution with mean $Q(s_k^t; \theta)$ and covariance matrix $Z$:
\begin{equation}
    q_t^k(u) \sim \mathbf{N}(Q(s_k^t; \theta), Z),
\end{equation}
where $Z$ is as described in Eq. \eqref{eq:ucb-bound}. Then the action with the highest estimated value is performed: 
\begin{equation}
    u^t_k = \argmax_{u} q_t^k(u).
\end{equation}
We also employ a linear version of this method, refereed to as ILin-TS, in which we use a linear function approximator to estimate the mean of the Gaussian posterior distribution used for estimating the action values.

\section{Evaluation}\label{sec:evaluation}
We evaluate our proposed framework through an extensive simulation.
In our simulations, we employ a full UHD $360^\circ$ video dataset \cite{Chakareski-2021-Full}.
This dataset includes 15 videos with various spatio-temporal characteristics.
Each video is divided into 36 segments of fixed time duration. Each video segment is a sequence of video frames. Seven layers of increased immersion fidelity for each frame is available.
As shown in Fig. \ref{fig:task-stats}, the perceived video quality by the users and task size increase as the number of layers included in each video segment increases. The increased task size introduces a higher computational requirements.
The elastic computational task that we consider here is to render each video segment.

Fig. \ref{fig:size-dist} shows the distribution of the size of elastic tasks.
In our simulation, we consider a multi-connectivity AR/VR system, where each user has access to three communication channels operating on 4G, 5G, and WiGig technologies.
We use a dataset of 4G and 5G network throughput traces~\cite{Narayanan-2021-Variegated}, which were collected in two different cities in the U.S. and from commercial operators (T-Mobile and Verizon).
We also employ a dataset of WiGig network throughput traces.
Fig. \ref{fig:datarate-cdf} shows the statistical characteristics (\ie cumulative distribution function (CDF), average, and standard deviation) of these network traces.
We employ these datasets to train all the agents mentioned above for $2,000$ episodes. 
To ensure a fair comparison, we use the same configuration for all the baselines as presented in Table \ref{tab:parameter-values}.
\begin{table}[ht]
    \centering
    \resizebox{.85\linewidth}{!}{
    \begin{tabular}{lc}
        \toprule
        Definition/Explanation & Parameter \& Value \\
        \midrule
        Number of VR users & $N=30$ \\
        Tasks deadline & $T_d^k = 1$ Second\\
        \midrule
        CPU capacitance factor & $\kappa = 1\mathrm{e}{-27}$ \\
        Users' CPU frequency & $\uavCPUFreq^k = 2.4 \,GHz$ \\
        User computing speed & $Z^{k}=100 \, Mbps$ \\
        MEC computing speed & $Z_{mec} = 2.5 \,Gbps$ \\
        \midrule
        5G transmission power & $5.27 \, mW/Mbps$ \\
        4G transmission power & $57.99 \, mW/Mbps$ \\
        WiGig transmission power & $6.15 \, mW/Mbps$ \\
        \midrule
        Number of training iterations & $N_{policy}=  80$, $N_{aux}= 6$, $N_{\lambda} = 5$ \\
        Initial deadline coefficient & $\lambda_k=16$ \\
        Policy update frequency & $N_{update}=4$ \\
        Entropy weight & $\beta=0.01$ \\
        \bottomrule
    \end{tabular}
    }
    \caption{Simulation and training parameter values.}
    \label{tab:parameter-values}
    \vspace{-.1in}
\end{table}

\begin{figure}
    \centering
\includegraphics[width=0.95\linewidth]{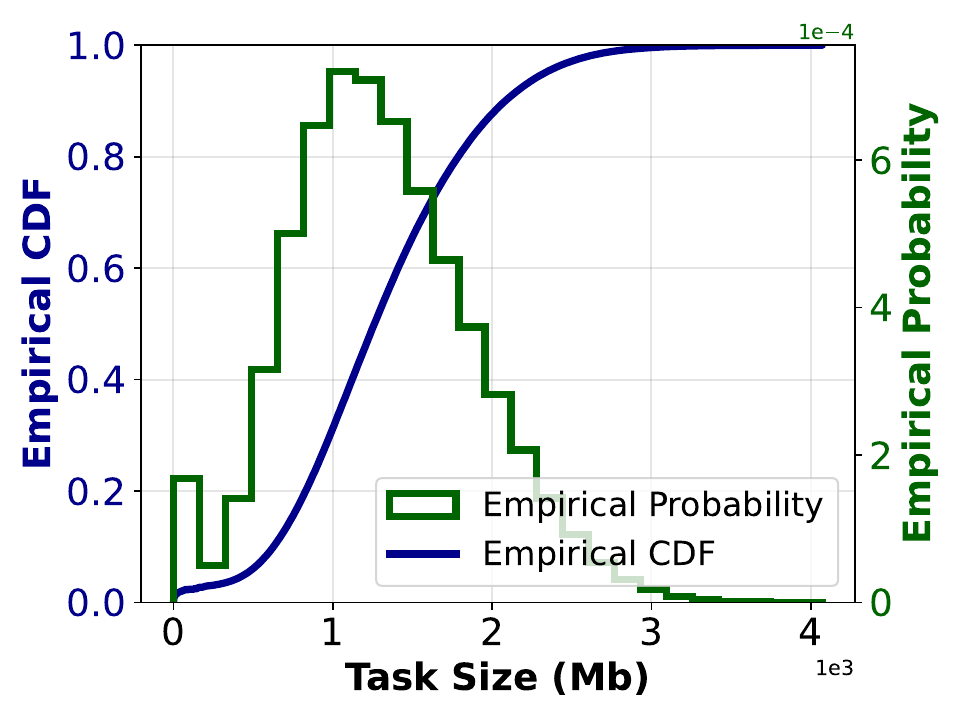}
    \caption{Empirical cumulative distribution function (CDF) and empirical probability of the elastic task size, taken from a real-world $360^\circ$ video dataset.}
    \label{fig:size-dist}
\end{figure}

\begin{figure}
    \centering
    \includegraphics[width=0.95\linewidth]{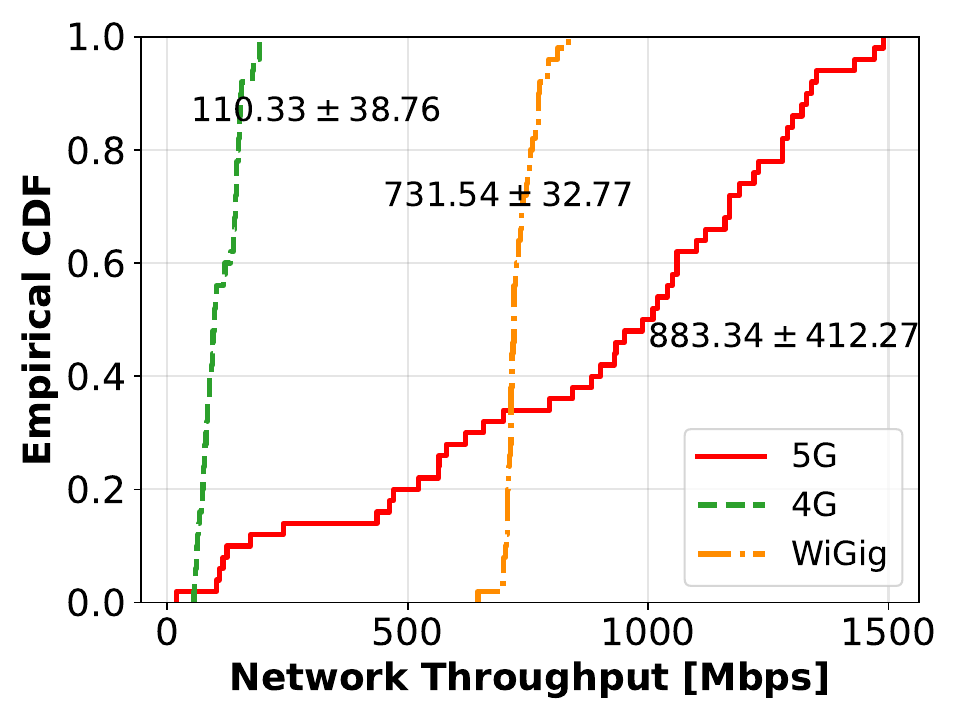}
    \caption{Network throughput cumulative distribution function (CDF) of 4G, 5G, and WiGig networks and their means and standard deviations. These statistics are collected from real-world network traces.}
    \label{fig:datarate-cdf}
\end{figure}

\textbf{Deployment Consideration:}
The proposed methods, CPPG, IPPG, and DSMAB variations, exhibit notable differences across several key aspects during training and deployment.
Table \ref{tab:baseline-comparision} summarizes these key differences. 
CPPG affords a full observability of the environment. Hence, CPPG utilizes a fully centralized approach for both training and deployment, granting a constant inference time complexity of $\mathcal{O}(1)$ to make an offloading decision for all $K$ users. However, CPPG requires all users' state information in order to make an offloading decision. 
IPPG, while maintaining centralized training, shifts to decentralized deployment. This, in turn, leads to partial observability and a linear inference time complexity of $\mathcal{O}(K)$ to make an offloading decision for all $K$ users.
Moreover, IPPG has a significantly reduced state size of $15$, which corresponds to size of communication and computation history and task information. This differs from the $15 \times K$ state size of CPPG.
DSMAB variants (\ie INN $\epsilon$-Greedy, INN-TS, INN-UCB,), also with centralized training and decentralized deployment, shares the $\mathcal{O}(K)$ inference time and partial observability with IPPG, but expands the state size to $15\times K \times C$ to accommodate contextual information of $C$ offloading actions.




\begin{table*}[h]
    \centering
    \begin{tabular}{|l|c|c|c|c|c|c|}
    \hline
         Algorithm & Training & Deployment & Observability & Inference Time Complexity & State Size & Exploration Strategy\\
    \hline
         CPPG & Centralized & Centralized & Full & $\mathcal{O}(1)$ & $15 \times K$ & Min. Entropy\\
         IPPG & Centralized & Decentralized & Partial & $\mathcal{O}(K)$ & $15$ & Min. Entropy\\
         DSMAB & Centralized & Decentralized & Partial & $\mathcal{O}(K)$ & $15 \times K \times C$ & UCB, TS, $\epsilon$-Greedy\\
    \hline
    \end{tabular}
    \caption{Comparison of the proposed methods in terms of training and deployment settings along with required inference time complexity and problem state space size. Centralized decision-making by CPPG achieves a lower inference time complexity at a expense of larger state space size and full observability during the deployment relative to the decentralized IPPG method.}
    \label{tab:baseline-comparision}
\end{table*}

\textbf{Deployment Performance:}
Fig. \ref{fig:deployment-performance} demonstrates the performance trade-offs between the task response time and energy consumption, which are collected from 1,000 episodes of testing.
Each point demonstrates the average task response time and task energy consumption. The vertical and horizontal bars represent the standard deviation of the response time and energy consumption, respectively.
Small response time and energy consumption with small variation are desirable, which is represented by a point in the lower left corner of these plots.
From the results, we observe that the CPPG provides the best performance trade-offs, and IPPG provides the second best performance trade-offs. 
This is due to the fact that the CPPG has access to the state of all agents, while IPPG has a partial observability of the environment (\ie only observes the state of each of the agent). This, in turn, enables the CPPG to better balance the trade-offs among agents.
However, this comes at the cost of higher computational complexity of the centralized decision-making, which is not a major concern due to the fact that the decision-making takes place on the MEC unit.

\begin{figure}[t]
    \centering
        \includegraphics[width=.95\linewidth]{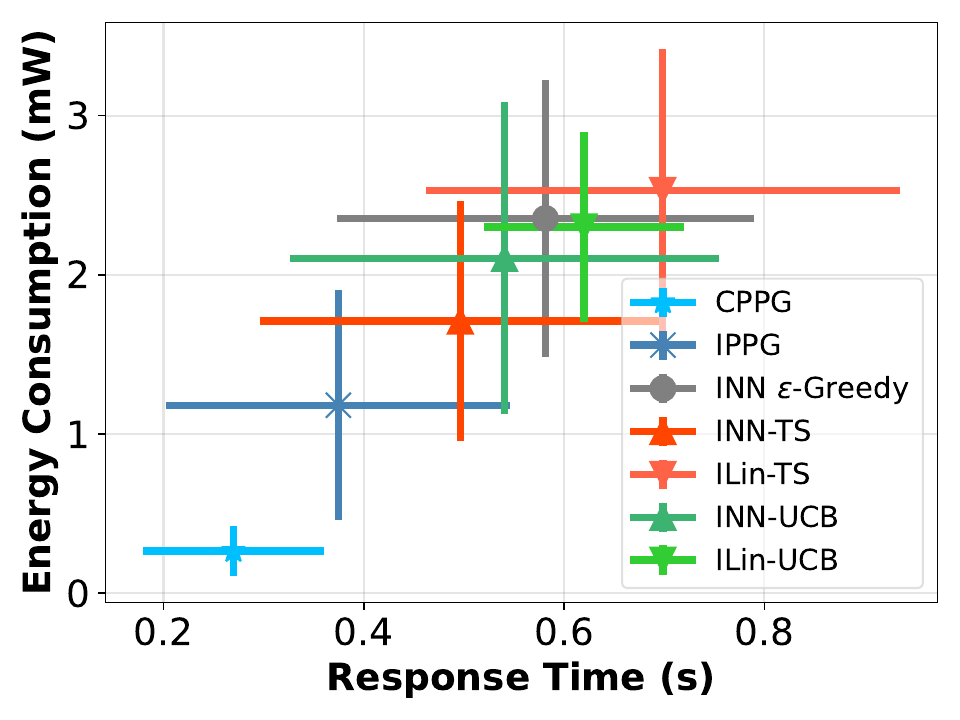}
    \caption{
    Performance trade-offs between response time and energy consumption in the testing stage for an environment with 30 users. 
    The CPPG provides the best performance trade-offs due to its full observability of the environment.
    }
    \label{fig:deployment-performance}
\end{figure}

\begin{figure*}[ht]
    \centering
    \begin{subfigure}{.305\linewidth}
        \centering
        \includegraphics[width=.99\linewidth]{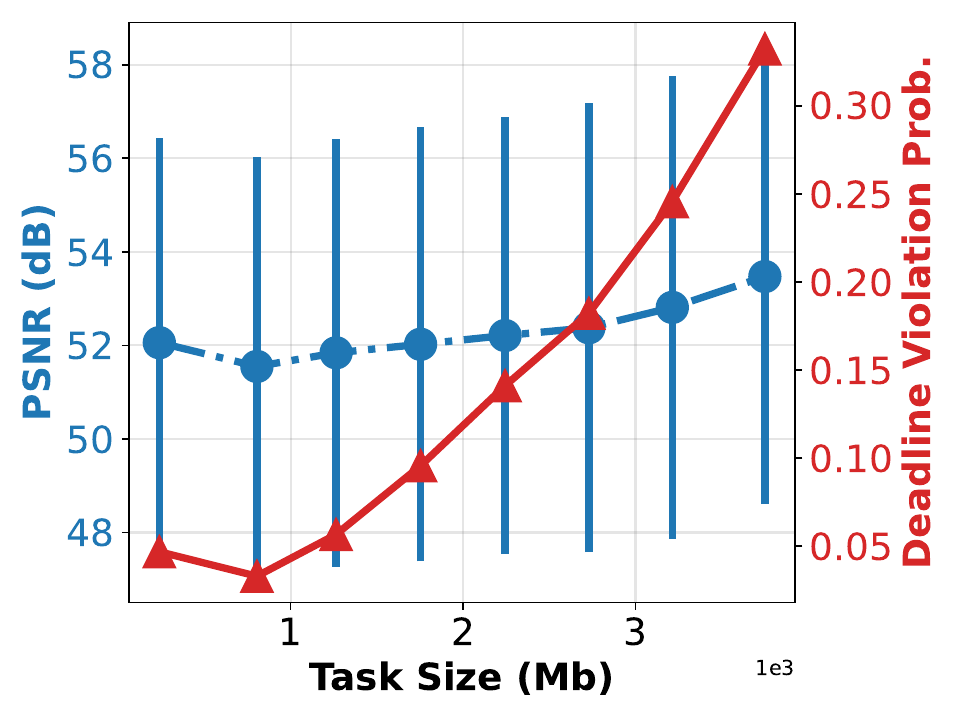}
        \caption{Quality of Experience Assessment}
        \label{fig:qoe-performance}
    \end{subfigure}
    \begin{subfigure}{.365\linewidth}
        \centering
        \includegraphics[width=.99\linewidth]{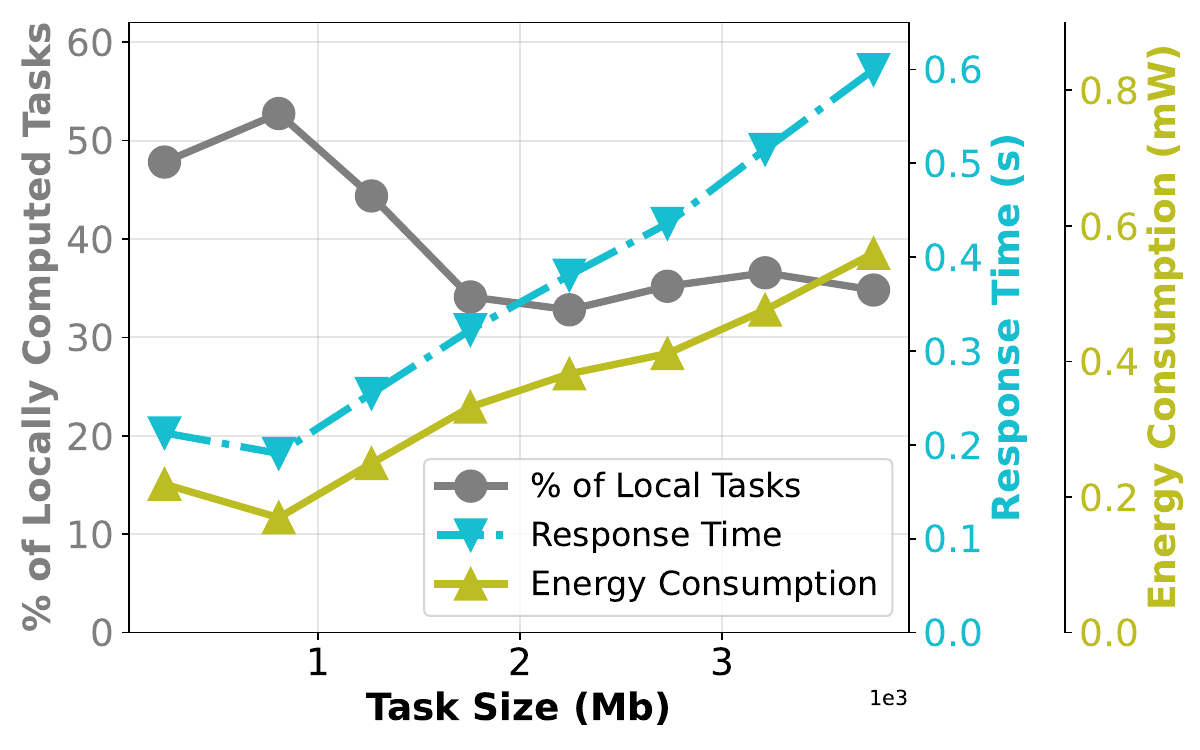}
        \caption{Task Size Effect Assessment}
        \label{fig:policy-analysis}
    \end{subfigure}
    \begin{subfigure}{.305\linewidth}
        \centering
        \includegraphics[width=.99\linewidth]{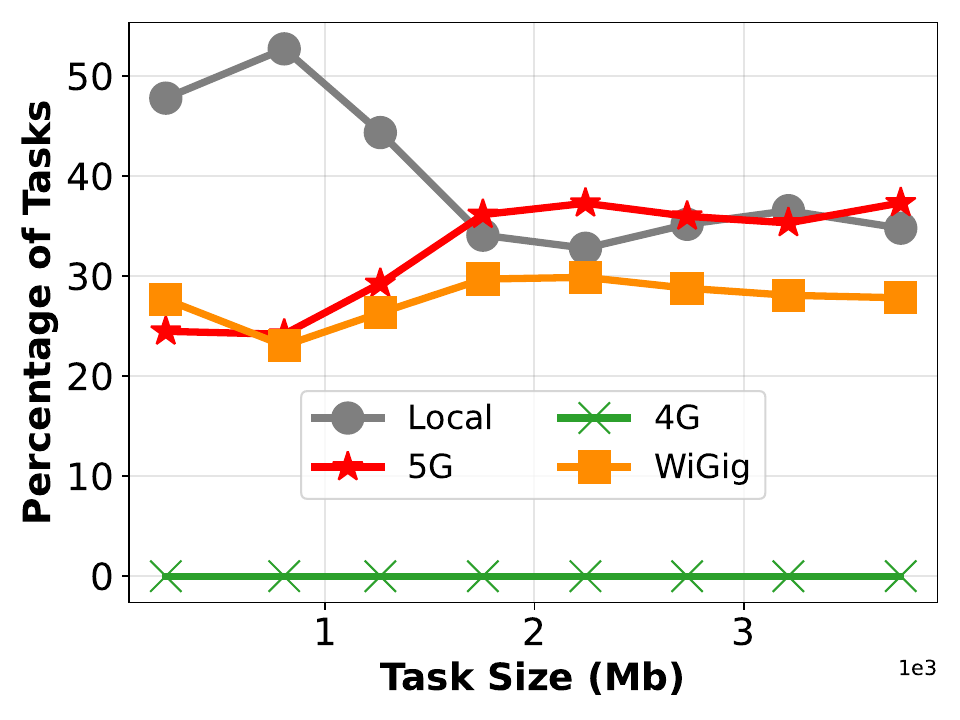}
        \caption{Offloading Policy Assessment}
        \label{fig:policy-detailed-analysis}
    \end{subfigure}
    \caption{
    a) The trade-offs between viewport PSNR and task deadline violations as task size increases.
    (b) Examination of task size impact on system performance and offloading strategy. As task size grows, CPPG reduces local computation, yet response time and energy consumption increase.
    (c) With increasing task size, CPPG favors 5G and WiGig channels due to their higher data rates, while avoiding the 4G channel because of its high energy consumption and lower data rate.
    }
    \label{fig:enter-label}
    \vspace{-.2in}
\end{figure*}
\textbf{Quality of Experience Assessment:}
Fig. \ref{fig:qoe-performance} demonstrates the performance trade-offs between viewport PSNR and task deadline violation. 
PSNR is a widely used metric that measures the quality of the video perceived by the end users in AR/VR applications.
Widespread use of PSNR as a objective metric of video quality is due to several factors. PSNR is easy to calculate, has a clear physical meaning, and is mathematically convenient for optimization~\cite{Zhou-2004-SSIM}.
PSNR joint with deadline violation probability models the quality of experience perceived by AR/VR users.
In our simulations, the deadline has been set in a way to capture the rebuffering events (\eg each elastic task is a video segment worth of one second, and the task deadline has been set to one second).
Fig. \ref{fig:qoe-performance} demonstrates that as the task size increases both PNSR and probability of violating the deadline increase. These observations align with elastic nature of tasks that was described in Fig. \ref{fig:task-stats}.
Note that the red curve shows the probability distribution of the tasks that have not met their deadline, which comprises 12.24\% of all tasks in the system. 
This means only 0.4\% of each user's tasks violated their deadlines in an environment with 30 users.



\textbf{Offloading Policy Assessment:}
Fig. \ref{fig:policy-analysis} demonstrates the CPPG's offloading policy given the task size and its effect on the system's performance (\ie the average response time and energy consumption).
These results are collected over 1000 episodes of testing in an environment with 30 VR users.
Fig. \ref{fig:policy-analysis} shows that the percentage of locally computed tasks decreases as the task size increases, which is due to higher computational demands of larger tasks.
Moreover, the response time and energy consumption increase as the task size increases, which is due to higher computational and communication delay.
This shows that CPPG is able to capture the effect of task size and task computational intensity and adjust its offloading policy based on the computational and communication requirements of the task.

Fig. \ref{fig:policy-detailed-analysis} provides deeper insights into the actions taken by the CPPG agent as the task size increases. Notably, the agent avoids offloading tasks via the 4G channel, as it offers the lowest network throughput and highest energy consumption compared to 5G and WiGig (see Fig. \ref{fig:datarate-cdf} and Table \ref{tab:parameter-values}). Additionally, the agent reduces the on-device computation on the VR headsets due to the higher computational demand of such tasks. Consequently, the offloading rate increases, with tasks distributed between the 5G and WiGig channels, where 5G handles a larger share due to its higher data rate.
Fig. \ref{fig:policy-detailed-analysis} further highlights the CPPG’s ability to effectively balance communication and computation trade-offs, ensuring an optimal experience for the VR users.






\textbf{Scalability Analysis:}
Fig. \ref{fig:scalability} demonstrates the performance of our proposed methods as the number of users in the environment increases. 
These results are collected from 1000 episodes of testing. 
In Fig. \ref{fig:scalability-a} and Fig. \ref{fig:scalability-b}, each bar shows the average task response time and energy consumption, respectively. The vertical lines represent the standard deviation of the response time and the energy consumption.
The response time and energy consumption are expected to grow as the the number of users in the environment increases since the MEC's computational resources is shared among all users.
However, we see that CPPG performance does not vary significantly as the number of users increases.
This is due to the fact that CPPG is a centralized decision making agent and has full observability of the environment, thus can capture the communication computation trade-offs in a multi-agent setting.
On the contrary, the other methods are decentralized and each agent makes an independent decision based on its partial observation of the state of the environment, thus each agent makes a sub-optimal decision.
\begin{figure*}[ht]
    \centering
    \begin{subfigure}{.48\linewidth}
        \centering
        \includegraphics[width=.99\linewidth]{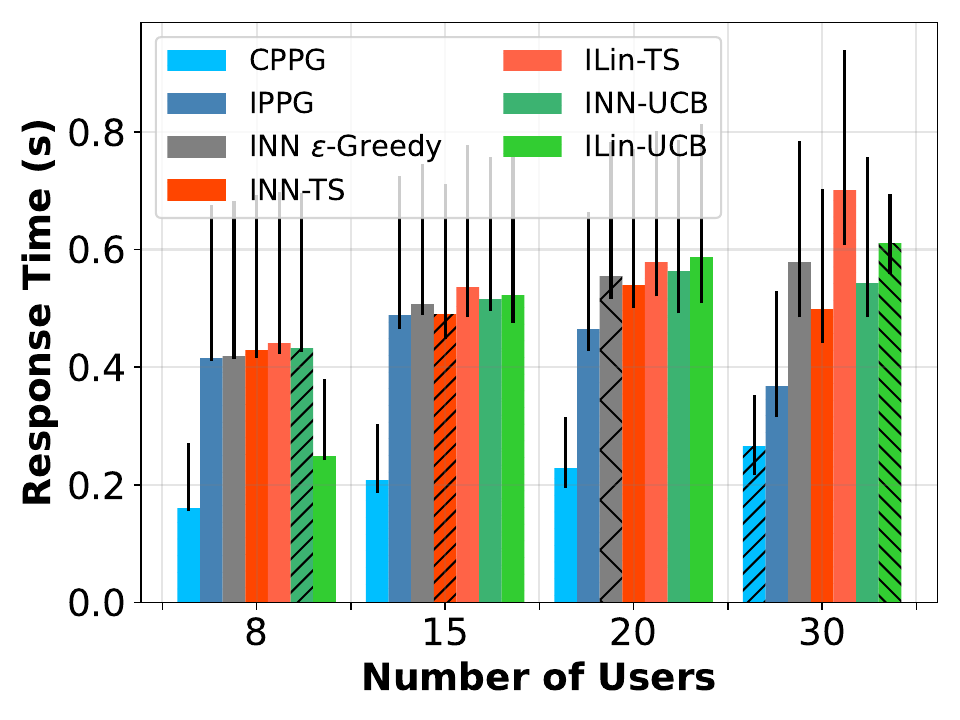}
        \caption{Response Time Scalability}
        \label{fig:scalability-a}
    \end{subfigure}
    \begin{subfigure}{.48\linewidth}
        \centering
        \includegraphics[width=.99\linewidth]{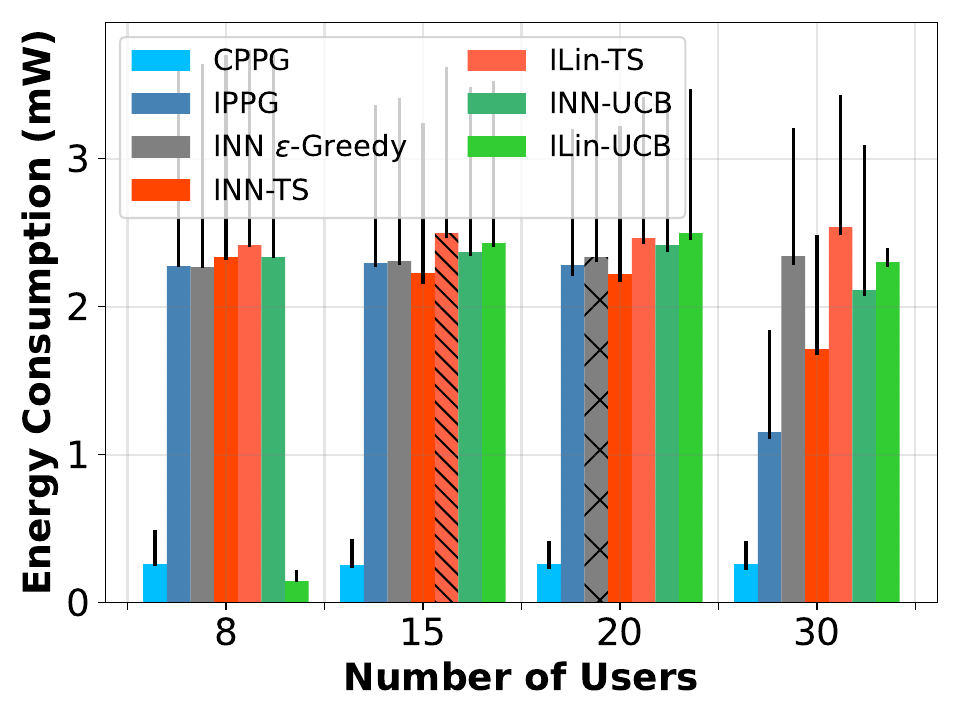}
        \caption{Energy Scalability}
        \label{fig:scalability-b}
    \end{subfigure}
    \caption{
    Scalability analysis of the proposed methods. Each bar and the vertical line represent the average and standard deviation of task response time and energy consumption, respectively. CPPG scales well in terms of performance due to its full observability of the environment at the expense of increasing state space size.
    }
    \label{fig:scalability}
    \vspace{-.2in}
\end{figure*}

\textbf{Computational Speed Effect Assessment:}
Fig. \ref{fig:comp-speed-analysis} investigates the effect of having extra computational resources on the VR headsets and the MEC unit on the system's performance (\ie the average response time and energy consumption) and the resulting offloading policy. In particular, Fig. \ref{fig:vr-comp-speed-analysis} reports the effect of increased computational resources on the VR headsets.
An increase in computational resources for the VR headset leads to an increase in local computation, thus a lower response time. 
This means that as the available computational resources on the headsets' increase, the CPPG method selects to compute more tasks locally. This, in turn, releases some of the computational resources on the MEC unit to offload more computationally intensive task to the MEC, which leads to a lower average response time for all of the tasks. 
However, an increase in locally computed tasks does not necessarily lead to higher energy consumption due to higher local CPU utilization. This is due to the fact that CPPG accounts for the computational energy-efficiency and balances the trade-offs between energy consumption and response time.

Finally, Fig. \ref{fig:mec-comp-speed-analysis} reports the effect of increased computational resources for the MEC unit.
An increase in computational resources of the MEC unit leads to a decrease in local computation and average response time due to having excess computational resources for computationally intensive tasks at the MEC.
However, the energy consumption slightly increases due to higher communication energy consumption required for transmitting the tasks over the air to the MEC.

\begin{figure*}
    \centering
    \begin{subfigure}{0.48\linewidth}
        \centering
        \includegraphics[width=.99\columnwidth]{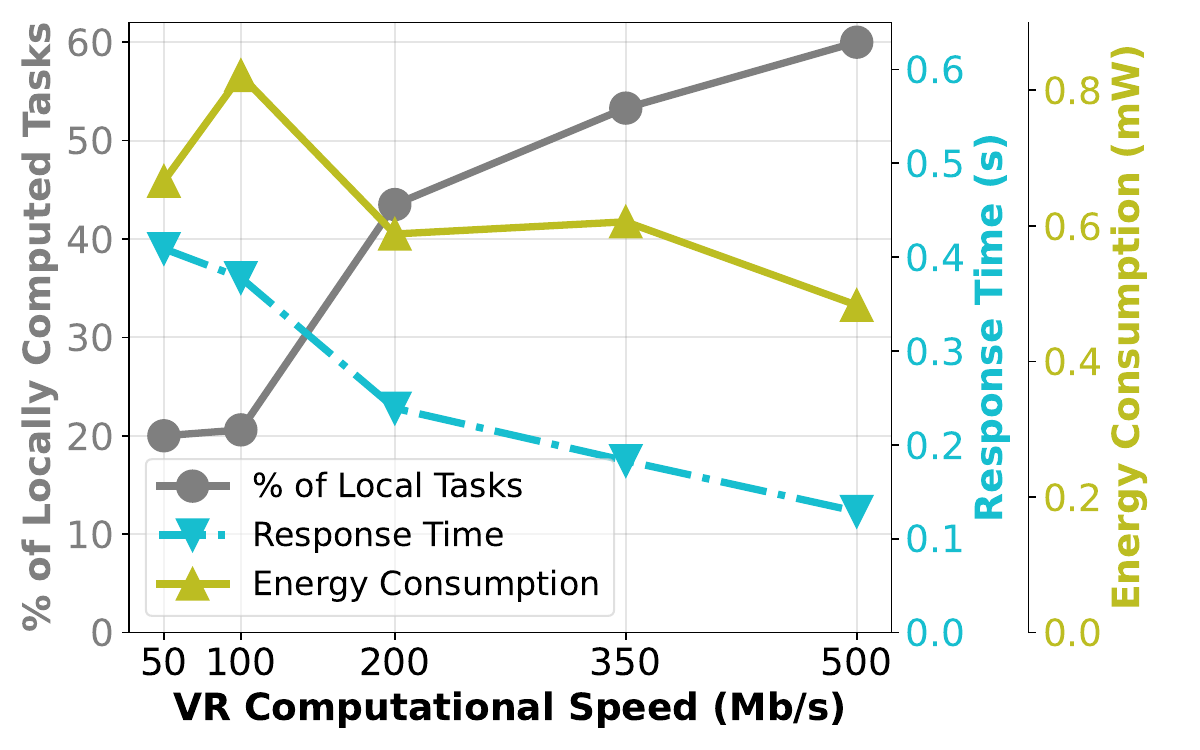}
        \subcaption{VR Computational Speed Analysis}
        \label{fig:vr-comp-speed-analysis}
    \end{subfigure}
    \begin{subfigure}{0.48\linewidth}
    \centering
        \includegraphics[width=.99\columnwidth]{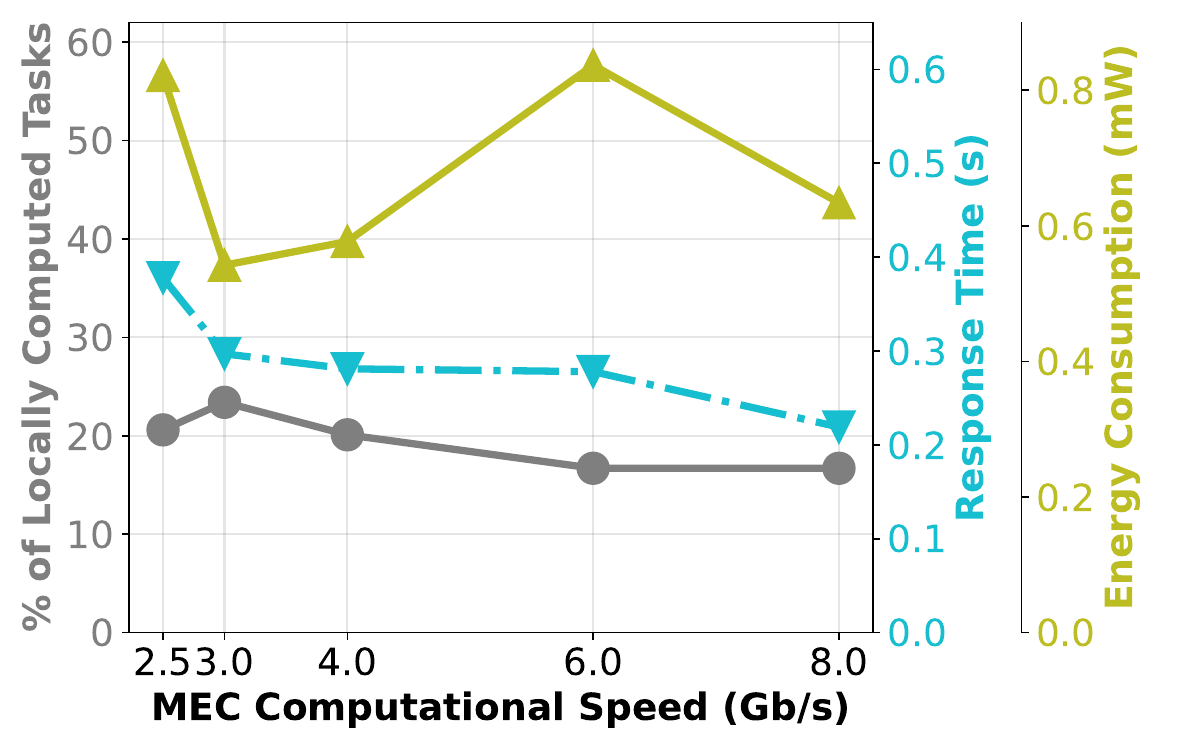}
        \subcaption{MEC Computational Speed Analysis}
        \label{fig:mec-comp-speed-analysis}
    \end{subfigure}
    \caption{
    (a) As the VR headset's computational speed increases, a higher percentage of tasks are computed locally, resulting in reduced response time and energy consumption. (b) As the MEC computational speed increases, a lower percentage of tasks are computed locally, leading to reduced response time. However, the energy consumption is not guaranteed to decrease as the percentages of offloaded elastic task increases, which leads to higher transmission energy consumption.
    }
    \label{fig:comp-speed-analysis}
    \vspace{-.2in}
\end{figure*}

\section{Conclusion}\label{sec:conclusion}
In this paper, we considered the problem of constrained stochastic multi-connectivity elastic task offloading in AR/VR applications.
We present a learning-based decision-making framework for elastic task offloading decision making. 
The overall objective is to optimize the computational energy-efficiency under time-varying conditions in terms of available computational resources, communication bandwidth, energy resources, and user requirements.
Leveraging a state-of-the-art DRL algorithm and MARL techniques, we develop CPPG, a centralized decision-making agent, that utilizes elastic task information (\ie size and computational intensity) and communication and computation history (\ie past transmission times, past response times, and past energy consumptions) to make an offloading decision.
Furthermore, leveraging the network sharing techniques, 
we extend our solution to a decentralized setting and present IPPG and DSMAB methods (\ie INN $\epsilon$-Greedy, INN-UCB, and INN-TS).
Through numerical simulation using real-world network traces and $360^\circ$ video information, we showed that the proposed methods learn to balance the existing trade-offs in the system and meet the task requirements (\ie task deadline).
In particular, CPPG and IPPG provide the best performance trade-offs. 
CPPG, in comparison to IPPG, reduces the latency by 28\% and energy consumption by 78\% at the cost of full observability of the environment.
Meanwhile, IPPG offers decentralized task offloading and lower computational complexity.
{\small
\bibliographystyle{IEEEtranN}
\bibliography{ref2}
}
\end{document}

%% file: notation.tex
\newcommand\task{T}

\newcommand\taskTxTime{\task_{tx}}

\newcommand\taskExecution{\task_{e}}

\newcommand\taskSize{S} 
\newcommand\taskIntencity{I} 
\newcommand\taskDeadline{\task_{d}} 
\newcommand\taskResponse{\task_{r}} 

\newcommand\resultSize{S_{res}} 
\newcommand\resultTxTime{\task_{rx}}

\newcommand\nbUAVs{K}
\newcommand\nbChannels{C}

\newcommand\cpuFreq{f} 
\newcommand\mecCPUFreq{\cpuFreq_{mec}} 
\newcommand\uavCPUFreq{\cpuFreq_{vr}} 

\newcommand\energy{E}
\newcommand\totalEnergy{\energy_{tot}}
\newcommand\cpuEnergy{\energy_{e}}
\newcommand\txEnergy{\energy_{tx}}
\newcommand\rxEnergy{\energy_{rx}}

\newcommand\indicator{\mathbbm{1}}

\newcommand{\ie}{{\textit{i.e., }}}
\newcommand{\eg}{{\textit{e.g., }}}

%% file: main.bbl
\begin{thebibliography}{62}
\providecommand{\natexlab}[1]{#1}
\providecommand{\url}[1]{#1}
\csname url@samestyle\endcsname
\providecommand{\newblock}{\relax}
\providecommand{\bibinfo}[2]{#2}
\providecommand{\BIBentrySTDinterwordspacing}{\spaceskip=0pt\relax}
\providecommand{\BIBentryALTinterwordstretchfactor}{4}
\providecommand{\BIBentryALTinterwordspacing}{\spaceskip=\fontdimen2\font plus
\BIBentryALTinterwordstretchfactor\fontdimen3\font minus \fontdimen4\font\relax}
\providecommand{\BIBforeignlanguage}[2]{{%
\expandafter\ifx\csname l@#1\endcsname\relax
\typeout{** WARNING: IEEEtranN.bst: No hyphenation pattern has been}%
\typeout{** loaded for the language `#1'. Using the pattern for}%
\typeout{** the default language instead.}%
\else
\language=\csname l@#1\endcsname
\fi
#2}}
\providecommand{\BIBdecl}{\relax}
\BIBdecl

\bibitem[SeaGate(2019)]{SeaGate-2019-State}
SeaGate, ``{State of the Edge},'' \url{https://www.seagate.com/www-content/enterprise-storage/it-4-0/images/Data-At-The-Edge-UP1.pdf}, 2019, [Online].

\bibitem[Chakareski et~al.(2022)Chakareski, Khan, and Yuksel]{ChakareskiKY:20}
J.~Chakareski, M.~Khan, and M.~Yuksel, ``{Towards Enabling Next Generation Societal Virtual Reality Applications for Virtual Human Teleportation},'' \emph{IEEE Signal Processing Magazine}, Sep. 2022.

\bibitem[Chakareski et~al.(2023)Chakareski, Khan, Ropitault, and Blandino]{Chakareski-2023-Millimeter}
J.~Chakareski, M.~Khan, T.~Ropitault, and S.~Blandino, ``{Millimeter Wave and Free-Space-Optics for Future Dual-Connectivity 6DOF Mobile Multi-User VR Streaming},'' \emph{ACM Trans. Multimedia Computing, Communications, and Applications.}, vol.~19, no.~2, pp. 1--25, Feb. 2023.

\bibitem[Insights()]{fortune-2024-vr}
F.~B. Insights, ``{Virtual Reality (VR) Market Size, Share \& Industry Analysis},'' \url{https://www.fortunebusinessinsights.com/industry-reports/virtual-reality-market-101378}, [Online].

\bibitem[Ghazikor et~al.(2024)Ghazikor, Roach, Cheung, and Hashemi]{ghazikor-2024-channel}
\BIBentryALTinterwordspacing
M.~Ghazikor, K.~Roach, K.~Cheung, and M.~Hashemi, ``Channel-aware distributed transmission control and video streaming in uav networks,'' 2024. [Online]. Available: \url{https://arxiv.org/abs/2408.01885}
\BIBentrySTDinterwordspacing

\bibitem[Cobbe et~al.(2020)Cobbe, Hilton, Klimov, and Schulman]{cobbe-2020-ppg}
K.~Cobbe, J.~Hilton, O.~Klimov, and J.~Schulman, ``{Phasic Policy Gradient},'' \emph{arXiv:2009.04416}, 2020.

\bibitem[Brockman et~al.(2016)Brockman, Cheung, Pettersson, Schneider, Schulman, Tang, and Zaremba]{brockman2016openai}
G.~Brockman, V.~Cheung, L.~Pettersson, J.~Schneider, J.~Schulman, J.~Tang, and W.~Zaremba, ``{Openai gym},'' \emph{arXiv:1606.01540}, 2016.

\bibitem[Ouyang et~al.(2019)Ouyang, Li, Chen, Zhou, and Tang]{Ouyang-2019-Adaptive}
T.~Ouyang, R.~Li, X.~Chen, Z.~Zhou, and X.~Tang, ``{Adaptive User-managed Service Placement for Mobile Edge Computing: An Online Learning Approach},'' in \emph{{IEEE INFOCOM}}, 2019.

\bibitem[Molina et~al.(2014)Molina, Muñoz, Pascual-Iserte, and Vidal]{Molina-2014-Joint}
M.~Molina, O.~Muñoz, A.~Pascual-Iserte, and J.~Vidal, ``{Joint scheduling of communication and computation resources in multiuser wireless application offloading},'' in \emph{{IEEE 25th PIMRC}}, 2014.

\bibitem[Wu et~al.(2021)Wu, Chen, Yang, and Wang]{Wu-2021-EdgeCentric}
B.~Wu, T.~Chen, K.~Yang, and X.~Wang, ``{Edge-Centric Bandit Learning for Task-Offloading Allocations in Multi-RAT Heterogeneous Networks},'' \emph{IEEE Tran. on Vehicular Technology}, 2021.

\bibitem[Wang et~al.(2022)Wang, Ye, and Lui]{Wang-2022-Decentralized}
X.~Wang, J.~Ye, and J.~C. Lui, ``{Decentralized Task Offloading in Edge Computing: A Multi-User Multi-Armed Bandit Approach},'' in \emph{{IEEE INFOCOM}}, 2022.

\bibitem[Li et~al.(2022)Li, Wang, Wu, and Jia]{Yang-2022-Optimal}
Y.~Li, T.~Wang, Y.~Wu, and W.~Jia, ``{Optimal dynamic spectrum allocation-assisted latency minimization for multiuser mobile edge computing},'' \emph{Digital Communications and Networks}, 2022.

\bibitem[Huang et~al.(2019)Huang, Li, Pan, Zheng, and Lu]{Huang-2019-Fine}
S.~Huang, L.~Li, Q.~Pan, W.~Zheng, and Z.~Lu, ``{Fine-Grained Task Offloading for UAV via MEC-Enabled Networks},'' in \emph{{IEEE 30th International Symposium on Personal, Indoor and Mobile Radio Communications (PIMRC Workshops)}}, 2019.

\bibitem[Jia et~al.(2021)Jia, Zhou, Wang, and Mumtaz]{Jia-2021-Learning}
Z.~Jia, Z.~Zhou, X.~Wang, and S.~Mumtaz, ``{Learning-Based Queuing Delay-Aware Task Offloading in Collaborative Vehicular Networks},'' in \emph{{IEEE ICC}}, 2021.

\bibitem[Liu et~al.(2020)Liu, Xu, Qi, Yao, Zhang, and He]{Liu-2020-Joint}
W.~Liu, Y.~Xu, N.~Qi, K.~Yao, Y.~Zhang, and W.~He, ``{Joint Computation Offloading and Resource Allocation in UAV Swarms with Multi-access Edge Computing},'' in \emph{{International Conference on Wireless Communications and Signal Processing}}, 2020.

\bibitem[Zhu et~al.(2018)Zhu, Gui, Chen, Zhang, and Zhang]{Zhu-2018-Cooperative}
S.~Zhu, L.~Gui, J.~Chen, Q.~Zhang, and N.~Zhang, ``{Cooperative Computation Offloading for UAVs: A Joint Radio and Computing Resource Allocation Approach},'' in \emph{{IEEE International Conference on Edge Computing (EDGE)}}, 2018.

\bibitem[Cao et~al.(2020)Cao, Ni, Tian, Hua, and Hao]{Cao-2020-UAV}
L.~Cao, W.~Ni, H.~Tian, M.~Hua, and G.~Hao, ``{UAV-Assisted Cellular System: Offloading Strategy and Bandwidth Allocation},'' in \emph{{International Conference on Space-Air-Ground Computing (SAGC)}}, 2020.

\bibitem[Zhang et~al.(2018)Zhang, Guo, and Liu]{Zhang-2018-Energy}
J.~Zhang, H.~Guo, and J.~Liu, ``{Energy-Aware Task Offloading for Ultra-Dense Edge Computing},'' in \emph{{IEEE iThings and IEEE GreenCom and IEEE CPSCom and IEEE SmartData}}, 2018.

\bibitem[Jiang et~al.(2023)Jiang, Dai, Xiao, and Iyengar]{Jiang-2023-Joint}
H.~Jiang, X.~Dai, Z.~Xiao, and A.~Iyengar, ``Joint task offloading and resource allocation for energy-constrained mobile edge computing,'' \emph{IEEE Transactions on Mobile Computing}, vol.~22, no.~7, pp. 4000--4015, 2023.

\bibitem[Yang et~al.(2022)Yang, Gao, Li, Qin, Sun, Zhang, Wang, and Li]{Yang-2022-Multi}
T.~Yang, S.~Gao, J.~Li, M.~Qin, X.~Sun, R.~Zhang, M.~Wang, and X.~Li, ``{Multi-Armed Bandits Learning for Task Offloading in Maritime Edge Intelligence Networks},'' \emph{IEEE Transactions on Vehicular Technology}, 2022.

\bibitem[Sacco et~al.(2021)Sacco, Esposito, Marchetto, and Montuschi]{Sacco-2021-Sustainable}
A.~Sacco, F.~Esposito, G.~Marchetto, and P.~Montuschi, ``{Sustainable Task Offloading in UAV Networks via Multi-Agent Reinforcement Learning},'' \emph{IEEE Trans. on Vehi. Tech.}, 2021.

\bibitem[Sacco et~al.(2022)Sacco, Esposito, Marchetto, and Montuschi]{Sacco-2022-Self}
------, ``{A Self-Learning Strategy for Task Offloading in UAV Networks},'' \emph{IEEE Transactions on Vehicular Technology}, 2022.

\bibitem[Chakareski et~al.(2020)Chakareski, Khan, Ropitault, and Blandino]{Chakareski-2020-6DOF}
J.~Chakareski, M.~Khan, T.~Ropitault, and S.~Blandino, ``{6DOF Virtual Reality Dataset and Performance Evaluation of Millimeter Wave vs. Free-Space-Optical Indoor Communications Systems for Lifelike Mobile VR Streaming},'' in \emph{Proc. 54\textsuperscript{th} Asilomar Conf. on Sig., Sys., and Computers}, 2020.

\bibitem[Hsu(2020)]{Hsu-2020-MEC}
C.-H. Hsu, ``{MEC-Assisted FoV-Aware and QoE-Driven Adaptive 360° Video Streaming for Virtual Reality},'' in \emph{16th Intern. Conf. on Mobility, Sensing and Networking (MSN)}, 2020.

\bibitem[Dai et~al.(2020)Dai, Zhang, Mao, and Liu]{Dai-2020-View}
J.~Dai, Z.~Zhang, S.~Mao, and D.~Liu, ``{A View Synthesis-Based 360° VR Caching System Over MEC-Enabled C-RAN},'' \emph{IEEE Transactions on Circuits and Systems for Video Technology}, 2020.

\bibitem[Maniotis and Thomos(2021)]{Maniotis-2021-Tile}
P.~Maniotis and N.~Thomos, ``{Tile-Based Edge Caching for 360° Live Video Streaming},'' \emph{IEEE Transactions on Circuits and Systems for Video Technology}, 2021.

\bibitem[Dang and Peng(2019)]{Dang-2019-Joint}
T.~Dang and M.~Peng, ``{Joint Radio Communication, Caching, and Computing Design for Mobile Virtual Reality Delivery in Fog Radio Access Networks},'' \emph{IEEE Journal on Selected Areas in Communications}, 2019.

\bibitem[Zhang and Chakareski(2022)]{Zhang-2022-UAV}
L.~Zhang and J.~Chakareski, ``{UAV-Assisted Edge Computing and Streaming for Wireless Virtual Reality: Analysis, Algorithm Design, and Performance Guarantees},'' \emph{IEEE Trans. on Vehicular Technology}, 2022.

\bibitem[Liu et~al.(2023)Liu, Wang, Huang, and Ye]{Liu-2023-On}
H.~Liu, S.~Wang, H.~Huang, and Q.~Ye, ``{On the Placement of Edge Servers in Mobile Edge Computing},'' in \emph{2023 International Conference on Computing, Networking and Communications (ICNC)}, 2023.

\bibitem[Guo et~al.(2021)Guo, Zhang, and Xia]{Guo-2021-Design}
Z.~Guo, P.~Zhang, and J.~Xia, ``{Design of Virtual Reality Education Platform based on 5G MEC},'' in \emph{2021 20th International Conference on Ubiquitous Computing and Communications}, 2021.

\bibitem[Aung et~al.(2024)Aung, Dhelim, Chen, Ning, Atzori, and Kechadi]{Aung-2024-Edge}
N.~Aung, S.~Dhelim, L.~Chen, H.~Ning, L.~Atzori, and T.~Kechadi, ``{Edge-Enabled Metaverse: The Convergence of Metaverse and Mobile Edge Computing},'' \emph{Tsinghua Science and Technology}, 2024.

\bibitem[Han et~al.(2019)Han, Lee, and Ok]{Han-2019-Real}
M.~Han, S.-H. Lee, and S.~Ok, ``{A Real-Time Architecture of 360-Degree Panoramic Video Streaming System},'' in \emph{2019 IEEE 2nd International Conference on Knowledge Innovation and Invention (ICKII)}, 2019.

\bibitem[Zhang et~al.(2023)Zhang, Wei, Wang, Ma, and Gao]{Zhang-2023-RealVR}
Q.~Zhang, J.~Wei, S.~Wang, S.~Ma, and W.~Gao, ``{RealVR: Efficient, Economical, and Quality-of- Experience-Driven VR Video System Based on MPEG OMAF},'' \emph{IEEE Transactions on Multimedia}, 2023.

\bibitem[Yang et~al.(2018)Yang, Chen, Li, Sun, Liu, Xie, and Zhao]{Yang-2018-Communication}
X.~Yang, Z.~Chen, K.~Li, Y.~Sun, N.~Liu, W.~Xie, and Y.~Zhao, ``{Communication-Constrained Mobile Edge Computing Systems for Wireless Virtual Reality: Scheduling and Tradeoff},'' \emph{IEEE Access}, 2018.

\bibitem[Lin et~al.(2021)Lin, Song, Wang, Yu, Guo, and Leung]{Lin-2021-Resource}
P.~Lin, Q.~Song, D.~Wang, F.~R. Yu, L.~Guo, and V.~C.~M. Leung, ``{Resource Management for Pervasive-Edge-Computing-Assisted Wireless VR Streaming in Industrial Internet of Things},'' \emph{IEEE Transactions on Industrial Informatics}, 2021.

\bibitem[Guo et~al.(2020)Guo, Yu, Zhang, Ji, Leung, and Li]{Guo-2020-AnAW}
F.~Guo, R.~Yu, H.~Zhang, H.~Ji, V.~C.~M. Leung, and X.~Li, ``{An Adaptive Wireless Virtual Reality Framework in Future Wireless Networks: A Distributed Learning Approach},'' \emph{IEEE Transactions on Vehicular Technology}, 2020.

\bibitem[Chen et~al.(2019)Chen, Saad, Yin, and Debbah]{Chen-2019-Data}
M.~Chen, W.~Saad, C.~Yin, and M.~Debbah, ``{Data Correlation-Aware Resource Management in Wireless Virtual Reality (VR): An Echo State Transfer Learning Approach},'' \emph{IEEE Trans. on Communications}, 2019.

\bibitem[Huang and Zhang(2018)]{Huang-2018-MAC}
M.~Huang and X.~Zhang, ``{MAC Scheduling for Multiuser Wireless Virtual Reality in 5G MIMO-OFDM Systems},'' in \emph{2018 IEEE International Conference on Communications Workshops (ICC Workshops)}, 2018.

\bibitem[Chakareski and Khan(2024)]{Chakareski-2024-Live}
J.~Chakareski and M.~Khan, ``{Live 360◦ Video Streaming to Heterogeneous Clients in 5G Networks},'' \emph{IEEE Transactions on Multimedia}, 2024.

\bibitem[Ge et~al.(2017)Ge, Pan, Li, Mao, and Tu]{Ge-2017-Multipath}
X.~Ge, L.~Pan, Q.~Li, G.~Mao, and S.~Tu, ``{Multipath Cooperative Communications Networks for Augmented and Virtual Reality Transmission},'' \emph{IEEE Transactions on Multimedia}, 2017.

\bibitem[Liu et~al.(2019)Liu, Liu, Argyriou, and Ci]{Liu-2019-MEC}
Y.~Liu, J.~Liu, A.~Argyriou, and S.~Ci, ``{MEC-Assisted Panoramic VR Video Streaming Over Millimeter Wave Mobile Networks},'' \emph{IEEE Transactions on Multimedia}, 2019.

\bibitem[Gupta et~al.(2023)Gupta, Chakareski, and Popovski]{Gupta-2023-mmWave}
S.~Gupta, J.~Chakareski, and P.~Popovski, ``{mmWave Networking and Edge Computing for Scalable 360$^\circ$ Video Multi-User Virtual Reality},'' \emph{IEEE Transactions on Image Processing}, 2023.

\bibitem[Ren et~al.(2019)Ren, He, Huang, Yu, Cai, and Zhang]{Ren-2019-Edge}
J.~Ren, Y.~He, G.~Huang, G.~Yu, Y.~Cai, and Z.~Zhang, ``{An Edge-Computing Based Arch. for Mobile Aug. Reality},'' \emph{IEEE Network}, 2019.

\bibitem[Younis et~al.(2020)Younis, Qiu, and Pompili]{Younis-2020-Latency}
A.~Younis, B.~Qiu, and D.~Pompili, ``Latency-aware hybrid edge cloud framework for mobile augmented reality applications,'' in \emph{2020 17th Annual IEEE International Conference on Sensing, Communication, and Networking (SECON)}, 2020, pp. 1--9.

\bibitem[Song and Shen(2023)]{Song-2023-Computing}
Y.~Song and Y.~Shen, ``{Computing Offloading Based on Deep Reinforcement Learning For Virtual Reality Scene},'' in \emph{2023 IEEE International Symposium on Broadband Multimedia Systems and Broadcasting (BMSB)}, 2023.

\bibitem[Cheng et~al.(2022)Cheng, Shan, Zhuang, Yu, Zhang, and S.~Quek]{Cheng-2022-Design}
Q.~Cheng, H.~Shan, W.~Zhuang, L.~Yu, Z.~Zhang, and T.~Q. S.~Quek, ``{Design and Analysis of MEC- and Proactive Caching-Based $360^\circ $ Mobile VR Video Streaming},'' \emph{IEEE Transactions on Multimedia}, 2022.

\bibitem[Didar and Brocanelli(2023)]{Didar-2023-eAR}
N.~Didar and M.~Brocanelli, ``ear: An edge-assisted and energy-efficient mobile augmented reality framework,'' \emph{IEEE Transactions on Mobile Computing}, vol.~22, no.~7, pp. 3898--3909, 2023.

\bibitem[Wang et~al.(2023)Wang, Kim, Xie, and Han]{Wang-2023-LEAF}
H.~Wang, B.~Kim, J.~Xie, and Z.~Han, ``Leaf + aio: Edge-assisted energy-aware object detection for mobile augmented reality,'' \emph{IEEE Transactions on Mobile Computing}, vol.~22, no.~10, pp. 5933--5948, 2023.

\bibitem[Nyamtiga et~al.(2022)Nyamtiga, Hermawan, Luckyarno, Kim, Jung, Kwak, and Yun]{Nyamtiga-2022-Edge}
B.~W. Nyamtiga, A.~A. Hermawan, Y.~F. Luckyarno, T.-W. Kim, D.-Y. Jung, J.~S. Kwak, and J.-H. Yun, ``{Edge-Computing-Assisted Virtual Reality Computation Offloading: An Empirical Study},'' \emph{IEEE Access}, 2022.

\bibitem[Zhang et~al.(2013)Zhang, Wen, Guan, Kilper, Luo, and Wu]{Zhang-2013-Energy}
W.~Zhang, Y.~Wen, K.~Guan, D.~Kilper, H.~Luo, and D.~O. Wu, ``{Energy-Optimal Mobile Cloud Computing under Stochastic Wireless Channel},'' \emph{IEEE Trans. on Wireless Comm.}, 2013.

\bibitem[Burd and Brodersen(1996)]{Burd-19960-Processor}
T.~D. Burd and R.~W. Brodersen, \emph{{Processor Design for Portable Systems}}, 1996.

\bibitem[Dusza et~al.(2013)Dusza, Ide, Cheng, and Wietfeld]{Dusza-2013-CoPoMo}
B.~Dusza, C.~Ide, L.~Cheng, and C.~Wietfeld, ``{CoPoMo: a context-aware power consumption model for LTE user equipment},'' \emph{Trans. on Emerging Telecomm. Tech.}, 2013.

\bibitem[Zappone and Jorswieck(2015)]{Zappone-2015-Energy}
A.~Zappone and E.~Jorswieck, \emph{{Energy Efficiency in Wireless Networks via Fractional Programming Theory}}, 2015.

\bibitem[Narayanan et~al.(2021)Narayanan, Zhang, Zhu, Hassan, Jin, Zhu, Zhang, Rybkin, Yang, Mao, Qian, and Zhang]{Narayanan-2021-Variegated}
A.~Narayanan, X.~Zhang, R.~Zhu, A.~Hassan, S.~Jin, X.~Zhu, X.~Zhang, D.~Rybkin, Z.~Yang, Z.~M. Mao, F.~Qian, and Z.-L. Zhang, ``{A Variegated Look at 5G in the Wild: Performance, Power, and QoE Implications},'' in \emph{{Proceedings of ACM SIGCOMM}}, 2021.

\bibitem[Ye et~al.(2020)Ye, Liu, Sun, Shi, Zhao, Wu, Yu, Yang, Wu, Guo, Chen, Yin, Zhang, Shi, Wang, Fu, Yang, and Huang]{Deheng-2020-Mastering}
D.~Ye, Z.~Liu, M.~Sun, B.~Shi, P.~Zhao, H.~Wu, H.~Yu, S.~Yang, X.~Wu, Q.~Guo, Q.~Chen, Y.~Yin, H.~Zhang, T.~Shi, L.~Wang, Q.~Fu, W.~Yang, and L.~Huang, ``{Mastering Complex Control in MOBA Games with Deep Reinforcement Learning},'' in \emph{{The 34 AAAI Conf. on AI}}, 2020.

\bibitem[Tianchi Huang ; Chao Zhou ; Rui-Xiao~Zhang(2023)]{Tianchi-2023-Buffer}
C.~W. L.~S. Tianchi Huang ; Chao Zhou ; Rui-Xiao~Zhang, ``{Buffer Awareness Neural Adaptive Video Streaming for Avoiding Extra Buffer Consumption},'' in \emph{{IEEE INFOCOM Conf.}}, 2023.

\bibitem[Schulman et~al.(2017)Schulman, Wolski, Dhariwal, Radford, and Klimov]{schulman-2017-proximal}
J.~Schulman, F.~Wolski, P.~Dhariwal, A.~Radford, and O.~Klimov, ``{Proximal Policy Optimization Algorithms},'' \emph{arXiv:1707.06347}, 2017.

\bibitem[Yu et~al.(2022)Yu, Velu, Vinitsky, Gao, Wang, Bayen, and Wu]{Yu-2022-PPO}
C.~Yu, A.~Velu, E.~Vinitsky, J.~Gao, Y.~Wang, A.~Bayen, and Y.~Wu, ``The surprising effectiveness of ppo in cooperative multi-agent games,'' in \emph{Proceedings of the 36th International Conference on Neural Information Processing Systems}, ser. NIPS '22.\hskip 1em plus 0.5em minus 0.4em\relax Red Hook, NY, USA: Curran Associates Inc., 2022.

\bibitem[Terry et~al.(2020)Terry, Grammel, Son, and Black]{Justin-2020-Parameter}
\BIBentryALTinterwordspacing
J.~K. Terry, N.~Grammel, S.~Son, and B.~Black, ``Parameter sharing for heterogeneous agents in multi-agent reinforcement learning,'' \emph{CoRR}, vol. abs/2005.13625, 2020. [Online]. Available: \url{https://arxiv.org/abs/2005.13625}
\BIBentrySTDinterwordspacing

\bibitem[Zhou et~al.(2020)Zhou, Li, and Gu]{Zhou-2020-Neural}
D.~Zhou, L.~Li, and Q.~Gu, ``{Neural Contextual Bandits with UCB-Based Exploration},'' in \emph{{Proceedings of 37th ICML}}, 2020.

\bibitem[Chakareski et~al.(2021)Chakareski, Aksu, Swaminathan, and Zink]{Chakareski-2021-Full}
J.~Chakareski, R.~Aksu, V.~Swaminathan, and M.~Zink, ``{Full UHD 360-Degree Video Dataset and Modeling of Rate-Distortion Characteristics and Head Movement Navigation},'' in \emph{ACM Multimedia Sys. Conf.}, 2021.

\bibitem[Wang et~al.(2004)Wang, Bovik, Sheikh, and Simoncelli]{Zhou-2004-SSIM}
Z.~Wang, A.~Bovik, H.~Sheikh, and E.~Simoncelli, ``Image quality assessment: from error visibility to structural similarity,'' \emph{IEEE Transactions on Image Processing}, vol.~13, no.~4, pp. 600--612, 2004.

\end{thebibliography}
